\documentclass[11pt,a4paper]{article}

\pdfoutput=1

\usepackage{jcappub}
\usepackage{amsmath,amssymb,epsf}
\usepackage{graphicx,color}
\usepackage{amsfonts}
\usepackage{hyperref}
\usepackage{color}

\def\g{\gamma}

\def\beqa{\begin{eqnarray}}
\def\eeqa{\end{eqnarray}}

\def\unit{\relax{\rm 1\kern-.26em I}}
\def\nada{\relax{\rm 0\kern-.30em l}}

\numberwithin{equation}{section}
\newcommand{\be}{\begin{equation}}
\newcommand{\ee}{\end{equation}}
\newcommand{\bea}{\begin{eqnarray}}
\newcommand{\eea}{\end{eqnarray}}
\newcommand{\barr}{\begin{array}}
\newcommand{\earr}{\end{array}}

\def\beq{\begin{equation}}
\def\eeq{\end{equation}}
\def\be{\begin{equation}}
\def\ee{\end{equation}}
\def\bea{\begin{eqnarray}}
\def\eea{\end{eqnarray}}

\vspace{-0.2cm}
\rightline{IFT-UAM/CSIC-15-042}

\DeclareRobustCommand{\SkipTocEntry}[4]{}

\title{Multifield Dynamics in Higgs-otic Inflation  }

 \author{S. Bielleman,}\author{L.E. Ib\'a\~nez,}\author{F.G. Pedro} 
\author{and I. Valenzuela}

 \affiliation{Departamento de F\'{\i}sica Te\'orica UAM and
Instituto de F\'{\i}sica Te\'orica UAM/CSIC,\\
Universidad Aut\'onoma de Madrid 
Cantoblanco, 28049 Madrid, Spain 
\\[8mm]}

\abstract{In Higgs-otic inflation a complex neutral scalar combination of the $h^0$ and $H^0$ MSSM Higgs fields plays the
role of inflaton in a chaotic fashion. The potential is protected from large trans-Planckian corrections 
at large inflaton if the system is embedded in string theory so that the Higgs fields parametrize a D-brane
position.  The inflaton potential is then given 
by  a DBI+CS D-brane action yielding an approximate linear behaviour at large field.
The inflaton scalar potential is a 2-field model with specific non-canonical kinetic terms.
Previous computations of the cosmological parameters (i.e. scalar and tensor perturbations) did not
take into account the full 2-field character of the model, ignoring in particular the
presence of isocurvature perturbations and their coupling to the adiabatic modes. It is well known that for generic 2-field potentials such effects may significantly alter the observational signatures  of a given model. We perform a full analysis of adiabatic and isocurvature 
perturbations in the Higgs-otic 2-field model. We show that the predictivity 
of the model is increased compared to the adiabatic approximation. Isocurvature perturbations moderately
feed back into adiabatic fluctuations. However, the isocurvature component is  exponentially damped by the end of inflation.
The tensor to scalar ratio varies in a region $r=0.08-0.12$, consistent with combined Planck/BICEP results. }

\begin{document}
\hfill {}

\maketitle

\section{Introduction}

It is very hard to protect fundamental scalars  in field theory  from large quantum corrections.  It is well known that these
quantum corrections easily drive the scalar masses to the size of the largest  ultraviolet cut-off in the theory.
Even so, light scalars seem to play an important role in our understanding of the observed properties of particle physics and
cosmology.  On the one hand the recent observation of the Higgs particle at the LHC \cite{Higgs}  is consistent with a fundamental
$SU(2)\times U(1)$ complex doublet remaining light well below any ultraviolet cut-off. On the cosmology side, 
precision data increasingly  favours the existence of an inflaton scalar, with mass well below the Planck scale, 
 leading to a period of fast expansion of the universe during which quantum perturbations generate CMB anisotropies and plant  the seeds for galaxy formation.

Low-energy supersymmetry provides a rationale for the stability of the Higgs mass against quantum fluctuations.
However, the fact that no trace of SUSY has been found yet at the LHC and the measured value of the Higgs mass
$m_h\simeq 125$ GeV cast some doubts on the idea of low-energy SUSY. Indeed, such a Higgs mass is quite high
compared  to the expectations from the MSSM and can only be reached for very heavy SUSY spectra.  
One can think of
disposing altogether with the idea of SUSY and admit that the Higgs field is fine-tuned to be light, perhaps on the basis of
anthropic arguments. Still this is probably not enough, since if we insist on the validity of  a non-SUSY SM structure 
at higher energies, the renormalisation group evolution of the Higgs self-coupling is such that the potential becomes
unstable (or metastable) for scales larger than $\sim 10^{10}$ GeV, possibly indicating the presence of new physics at
these scales, well below the Planck scale (see e.g.\cite{elias}).

A most elegant way to stabilise the Higgs potential for scales above $\sim 10^{10}$ GeV is again SUSY. Indeed, the scalar potential of
e.g. the MSSM is positive definite for scales above SUSY breaking and hence the potential is automatically stable.
It was shown in \cite{imrv,Ibanez:2013gf} (see also \cite{hebecker1,hebecker2,otrosinter}) that under very general assumptions, if SUSY is broken at scales $10^{10-13}$ GeV, the resulting 
Higgs mass at low-energies is consistent with its measured mass $\sim 125$ GeV. Of course, with  such a high SUSY breaking scale 
the Higgs mass is again unprotected and one should admit fine-tuning as the ultimate reason for its lightness.
This fine-tuning could come from an underlying multiverse within string theory, which would also be at the root of the
understanding of the smallness of the cosmological constant. 

On the cosmology side, a fundamental scalar degree of freedom, the inflaton,  seems to provide the simplest explanation for
a variety of cosmological observations. That scalar again will have to be protected from acquiring a large mass
and interactions that spoil slow-roll.  
On the other hand, as we already remarked,  light scalars usually do not stay light  in a field theory.  Intuitively,  one may  expect having two 
fundamental scalars light to be even more unlikely than a single (Higgs) one. 
Thus if the Higgs field itself
could act as the inflaton the combination  Higgs/inflaton would look  more {\it likely } within a   landscape of theories. Of course,
independently from any landscape argument, it is a natural question whether the Higgs and inflaton scalar could be one and the same.
This has been considered often in the past, see e.g.\cite{Shaposhnikov:2009pv,Bezrukov:2012sa,Bezrukov:2012hx,Bezrukov:2014bra,Barbon:2009ya,Barbon:2015fla,Ellis:2014dxa,Terada:2015cna}, see \cite{higgsflation} for a review and references.

In ref. \cite{Ibanez:2014kia} it was proposed that the neutral Higgs system of the MSSM with SUSY broken at a large scale $\sim 10^{13}$ GeV 
could be in charge of cosmic inflation. This goes under the name of
Higgs-otic inflation  \cite{higgsotic}, since it is in some sense a MSSM Higgs version of Linde's chaotic
inflation \cite{chaotic}. This proposal is quite economical since it addresses several issues simultaneously. 
It provides stability for the Higgs scalar potential at the right scale, is consistent with the observed value of the Higgs mass
and a neutral Higgs component acts as a complex inflaton.  The inflaton has a trans-Planckian field range and leads to a certain variety 
of (non-canonical) 2-field  chaotic like inflation.  In order for this to be sensible  we need a theory of quantum gravity, which is naturally 
identified with  string theory.
The Higgs/inflaton vev corresponds to the position of a D-brane moving on periodic cycles in a version of {\it monodromy inflation}
\cite{Silverstein:2008sg,McAllister:2008hb,peloso,Gur-Ari:2013sba,Berg:2009tg,Palti:2014kza}, \cite{Marchesano:2014mla},\cite{Blumenhagen:2014gta,Franco:2014hsa,McAllister:2014mpa,Blumenhagen:2014nba,Hayashi:2014aua,Hebecker:2014eua,Arends:2014qca,Hebecker:2014kva,Garcia-Etxebarria:2014wla,Blumenhagen:2015qda,Ben-Dayan:2014lca},
\cite{Kaloper:2008fb,Kaloper:2011jz,Kaloper:2014zba}.
The potential for the Higgs/inflaton is dictated by the Dirac-Born-Infeld  + Chern-Simon action of the moving D-brane. Fluxes induce
a scalar quadratic potential for the Higgs/inflaton system, with masses of order the SUSY breaking scale $M_{SS}$. However, the kinetic term of the scalars is not
canonical but rather is determined by the scalar potential itself. This leads to a {\it  flattening } of the scalar potential at large fields \cite{Dong:2010in},
with a leading linear behaviour at large inflaton. In general, a substantial contribution of tensor perturbations is generated. 

The Higgs-otic model is a 2-field inflaton system.
In \cite{higgsotic}  a study of the cosmological observables was made for the model focusing only on the adiabatic perturbations and
ignoring possible 2-field specific effects like the generation of isocurvature perturbations. This was an important pending issue since
2-field effects can, in principle,  substantially modify the cosmological observables and furthermore the Planck satellite has provided stringent
bounds on isocurvature perturbations. In the present paper we perform a systematic analysis  of the observables  in  the Higgs-otic 2-field inflation system.
We find that, as expected, adiabatic and isocurvature perturbations form a coupled system and there is super-horizon evolution of 
the curvature  perturbations. This leads in general to a relative increase of adiabatic perturbations and consequently to a reduction of the 
tensor to scalar ratio $r$ compared to the computation in ref.  \cite{higgsotic}. The range of variation of $n_s$ gets smaller and is centered around the region
alowed by Planck data with a tensor to scalar ratio in a range $r=0.08-0.12$. Moreover, the isocurvature component is always very suppressed at the end of inflation, consistent with upper Planck bounds.

The structure of this paper is as follows. We review the main points of Higgs-otic inflation in the next section, in which the relevant definitions 
and the  inflaton potential are described. In section \ref{sec:2field} we review the main issues of 2-field inflation as  applied to the case of Higgs-otic inflation. 
Section \ref{sec:results}  presents the results for Higgs-otic inflation for three representative points in the parameter space of the induced soft-terms for the
Higgs-inflaton system. The latter are determined by a real positive parameter $0\leq A\leq 1$, with $A=(m_H^2-m_h^2)/(m_H^2+m_h^2)$,
$H,h$ being the neutral Higgs scalars driving inflation \cite{higgsotic}.
The first case  ($A= 0.83$) correspond to the {\it canonical} Higgs-otic model in which the lightest scalar field at the minimum of the potential 
(at  scale $M_{SS}$)  can be identified with the  SM Higgs field.  A second case ($A=0.7$) analyses  how those results are changed if there is
some effect (like modified RG running) slightly modifying the Higgs-otic setting. For completeness we finally  present a third case with $A=0.2$
in which the inflaton cannot be identified with the MSSM Higgs fields but could be relevant in some extensions of the MSSM.
At the end of this section we show the expectations for the $r-n_s$ plot in the Higgs-otic model for a variety of initial conditions and
mass scales.  Finally, section \ref{sec:conclusions} is left for the conclusions.

\section{Higgs-otic Inflation}\label{sec:Higgsotic}

In this section we briefly review and extend the results presented in \cite{Ibanez:2014kia,higgsotic} on Higgs-otic inflation. 
We thus motivate the form of the non-minimal 2-field inflaton potential to be analysed in the following sections.
Readers interested only in the inflationary analysis may safely jump to the next section.

Higgs-otic inflation refers to theories in which the inflaton is a complex scalar giving rise to gauge symmetry
breaking,  while attaining large field inflation.  The most obvious and natural candidate for that is the SM Higgs field itself, as described 
in \cite{higgsotic} .  Nevertheless the same idea may be applied to other  BSM  fields introduced for other purposes,
as we briefly discuss below. 
The essential ingredient 
is the identification of a complex inflaton with the position moduli of some $Dp$-brane system in string compactifications.
\footnote{The same structure appears in  terms of continuous complex Wilson lines in models with extra dimensions.
From the string theory point of view they correspond to equivalent T-dual compactifications.}
The motion of the brane corresponds to the gauge symmetry breaking through a scalar vev.
For concreteness we will consider the position moduli of a $D7$-brane describing cycles in a $T^2$ torus inside a 
Type IIB orientifold compactification, although the setting may be easily generalized to other string configurations
\footnote{For an introduction to orientifold constructions see \cite{BOOK} and references therein.}. 
As we emphasised above,
the reason to go to a string setting in order to implement this idea is twofold. First, we will be interested in producing 
a large field model in which inflation implies trans-Planckian excursions of the inflaton.  In order to do so in a
sensible manner we need a theory of quantum gravity. Indeed within string theory the vevs of scalar fields may be
trans-Planckian and one can still maintain an effective  potential Lagrangian which makes sense. Second,
string theory has the required properties/symmetries in order to keep under control Planck-supressed corrections
which become obviously important at trans-Planckian field values.  There are modular and shift symmetries which force
these potential corrections to be subleading. 

So let us consider,  for definiteness, a Type IIB orientifold yielding a theory with $N=1$ supersymmetry in $D=4$.
Gauge interactions appear from stacks of $D7$-branes wrapping 4-cycles in the CY compact manifold. Those will include 
the SM gauge interactions as well as the SM fields. They come from zero modes of   $D=8$ complex scalars $\Phi$ and $A$
upon reduction to $D=4$. In particular, the position of the $D7$-branes is parametrized by the vevs of $\Phi$ zero modes.
On a stack of $N$ $D7$ the gauge group is $U(N)$ and the scalars transform as adjoint chiral multiplets.  The gauge symmetry
may be reduced to that of the SM (or some extension)  by e.g. an orbifold twist in the compactification. Generically 
only some fields of the adjoint scalars $\Phi$ survive,  like those parametrising the motion of those D7-branes which can 
move along flat directions.   This is the case of $D7$-branes associated to SM Higgs vevs in ref.\cite{higgsotic}. 

\begin{figure}[ht]
\centering
{\includegraphics[width=1\textwidth]{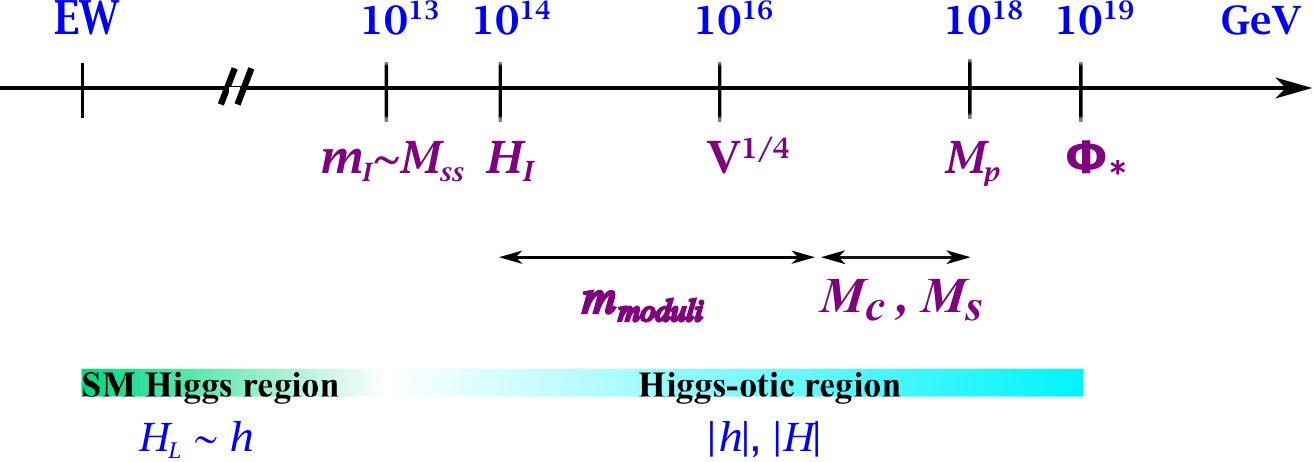}}
\caption{\small{Energy scales in the Higgs-otic Inflation scenario \cite{higgsotic}. Below $10^{13}$ GeV the light degrees of freedom in the Higgs sector are given by the $SU(2)$ doublet $H_L$. Above this scale $SU(2)$ is broken and they lie within the neutral components of $h$ and $H$.}}
\label{scalesfig}
\end{figure}

The generic presence of closed string fluxes in Type IIB compactifications  will give rise to a non-trivial potential for these fields, see ref.\cite{Camara:2004jj,Marchesano:2004yn,Giddings:2001yu,Camara:2003ku,Grana:2003ek,Lust:2004fi,Aparicio:2008wh,Camara:2014tba}. 
We thus assume, as is customary, that there are imaginary self-dual (ISD) fluxes $G_3$ acting as a background.  Such backgrounds are
known to be solutions of Type IIB $D=10$ equations of motion in warped CY backgrounds. In these backgrounds there are two classes
of ISD fluxes, with tensor structure  $G_{(0,3)}$ and $G_{(2,1)}$, respectively.  The first correspond locally to
components $G_{{\bar 1}{\bar 2}{\bar 3}}$ and the second to $G_{{\bar 1}{\bar 2}3}$. From now on we will denote
$G\equiv G_{{\bar 1}{\bar 2}{\bar 3}}$ and $S\equiv \epsilon_{3jk}G_{3{\bar j}{\bar k}}$.  The first class of flux, $G$,  breaks SUSY and
gives rise to SUSY-breaking soft terms, scalar and gaugino masses. The second  class, $S$, preserves SUSY and gives rise to
supersymmetric F-terms, i.e., $\mu$-terms.   These induced masses will eventually be identified with the mass scale of the inflaton. The cosmological bounds on the density scalar perturbations applied to our model fix this scale to be of order $10^{12}-10^{13}$ GeV \cite{Ibanez:2014zsa}.  In order to identify the inflaton with an MSSM Higgs boson we will see that the simultaneous presence of both classes of fluxes, G and S, is required.
This implies that the supersymmetry breaking scale $M_{ss}$ (ie. the scale of the soft terms)
 in this large field inflation setting will also be of order $10^{12}-10^{13}$ GeV. This is consistent with the scenario of Intermediate SUSY breaking in
 \cite{otrosinter,hebecker1,imrv,hebecker2}
which was shown to be consistent with a Higgs mass of $126$ GeV \cite{Ibanez:2013gf}.
The structure of mass scales in the Higgs-otic setting is summarised in fig.(\ref{scalesfig}).

 The dynamics of the fields living on the $D7$-branes is described by the Dirac-Born-Infeld (DBI) 
plus the Chern-Simons (CS) actions \cite{BOOK}.  Inserting the above two classes of flux backgrounds $G,S$ in the DBI+CS actions one obtains an effective $D=4$ bosonic action \cite{higgsotic}
\beq
\mathcal{L}_{4d}=\text{STr} \left(f(\Phi)D_\mu\Phi D^\mu\bar\Phi+\frac{1}{4g_{YM}^2}F_{\mu\nu}F^{\mu\nu}-V_F(\Phi)-\frac12 g_{YM}^2[\Phi,\bar \Phi]^2+...\right), 
\label{L4d}
\eeq
where there is a non-canonical kinetic factor $f(\Phi)$ which depends on the scalar
(F-term) potential $V_F$ as
\beq
f(\Phi)=1\ +\ \frac {(V_4\mu_7g_s)^{-1}}{2}V_F \ ,
\eeq
with
\beq
V_F(\Phi)=\frac{Z^{-2}g_s}{2}\ \text{Tr}\  |G^*\Phi-S\bar\Phi|^2\ \ .
\label{primpot}
\eeq
and Str denotes the symmetrised trace over gauge indices.
Here $g_s$ is the string coupling, $V_4$ the volume wrapped by the $D7$-brane and $Z^{-2}=1$ in the absence of warping, which we will assume 
in what follows. In addition there is the $D7$ tension $\mu_7= M_s^8/(2\pi)^7$, with $M_s= (\alpha')^{-1/2}$ the string scale.
To obtain Eq. \eqref{L4d} we have kept all the terms involving higher order powers of the position modulus $\Phi$, as required in a consistent large field inflation model. This is doable because the DBI action contains all perturbative corrections in $\alpha'$, whose effect is encoded in the non-canonical kinetic factor $f(\Phi)$. For small field $f\sim 1$ and we recover the result prescribed by the 4d supergravity effective action \cite{higgsotic}.

The above action concerns the $U(N)$ adjoint in the world volume of $D7$-branes. However, it still applies after
we make an orbifold projection that converts the adjoint into a set of bifundamentals, some of which are identified with the
Higgs field. For example \cite{Ibanez:2014kia,higgsotic}  one can start from a set of 6 $D7$-branes yielding an initial $U(6)$ gauge and
project out yielding a $U(3)\times U(2)\times U(1)$ gauge group. One linear combination of the $U(1)$'s can be identified
with hypercharge, whereas the other two are anomalous and become massive in the usual way. The adjoint $\Phi$ contains
doublets surviving projection, i.e.
\begin{equation}
\Phi=\left(\begin{array}{ccc}{\bf 0}_3&&\\& {\bf 0}_2&H_u\\&H_d&0\end{array}\right) \ , 
\label{phi}
\end{equation}
where $H_u$ and $H_d$ can be identified with the usual MSSM Higgs fields.
Plugging this decomposition in Eq. (\ref{primpot}) and taking the trace one obtains
the scalar potential 
\footnote{The same scalar potential may be obtained in an $N=1$ supergravity formulation 
not resting on the DBI+CS action. However, such a SUGRA treatment will miss the important
effect of the non-canonical kinetic term, which gives rise to the flattening of the potential, see \cite{higgsotic}.}
\begin{equation}
V_F=\frac{Z^{-2}g_s}{2}\left[(|G| -|S|)^2|h|^2+(|G|+|S|)^2|H|^2\right] \ ,
\label{VhH2}
\end{equation}
where one defines
\beq
h\, =\, \frac {e^{i\g/2}H_u - e^{-i\g/2}H_d^*}{\sqrt{2}}  \quad  ,  \quad H\, =\, \frac {e^{i\g/2}H_u+e^{-i\g/2}H_d^*}{\sqrt{2}} \ ,
\label{hH}
\eeq
with  $\g = \pi - {\rm Arg} (GS)$ given by the relative phase of both fluxes. 

The model is essentially double chaotic inflation with the key difference of non-canonical kinetic terms.
To see different limits of this expression it
is useful to define the  real variable $A$ which controls the relative size of SUSY (S) versus non-SUSY (G) fluxes, i.e.
\begin{equation}
A=\frac{2|SG|}{|G|^2+|S|^2}  \ .
\label{defA}
\end{equation}
Note that $0\leq A\leq 1$.  This parameter may also written in terms of the masses of the above defined
neutral scalars,
\beq
\frac {m_H}{m_h}\ =\ \sqrt{ \frac {1+A}{1-A}} \ .
\label{Amasses}
\eeq
There are two interesting limits in which this 2-field scalar potential becomes
effectively a single field potential:
\begin{itemize}
\item $A=1$. In this case $|G|=|S|$ and the field $h$ becomes massless. 
 If we want to eventually fine-tune a
Higgs doublet to remain at low-energies as a SM Higgs, much below the inflaton mass scale, 
we would need to be close to that situation. The Higgs field $H$ will be the one producing
inflation in this limit.

\item   $A=0$. This happens, in particular,  in the  SUSY case in which $G=0$. In this case 
the inflaton mass would be supersymmetric and $\sigma=\frac {1}{\sqrt {2}}\sqrt{|h|^2+|H|^2}$ 
acts as an effective single inflaton.  In this case one can hardly identify any of these scalars 
with (neutral) SM Higgs fields,  since they will have large masses, of order the inflaton mass
$\simeq 10^{13}$ GeV. On the other hand such fields could be identified with other BSM fields
which were not required to survive at low energies and could act as inflatons. Examples of these could be
$SU(5)$ scalar triplets or  $SU(2)_R$ doublets of left-right symmetric models.

\end{itemize}

The most economic option is of course identifying the fields with the MSSM Higgs doublets.
In such a case we would need one full  SM Higgs doublet to remain massless below
the SUSY breaking scale $\simeq 10^{13}$ GeV. This requires fine-tuning of the fluxes
$G,S$. However, one has to take into account the running of masses in between the  UV scale (the compactification/string scale $M_s$)
and the SUSY breaking scale $M_{ss}\simeq 10^{13}$ GeV. Taking into account this effect it was found in 
ref(...) that  one needs $A\simeq 0.83$.  Despite this fact,  in our inflationary analysis we will present results for a range of possible $A$ values.  It should be kept in mind that only results in which $A$ is not far from a central value
$A\simeq 0.83$ are consistent with the identification of the inflaton with MSSM Higgs fields.  Other values of $A$ could correspond 
to other BSM alternatives, as discussed above.

It is useful to write down the scalar potential in Eq. (\ref{VhH2}) in angular variables. Defining the D-term flat direction
as
\beq
\sigma=|H_u|=|H_d| \quad , \quad \quad H_u =   e^{i\theta} H_d^* \ ,
\label{sigthetadef}
\eeq
one rewrites the potential as
\begin{equation}
V(\sigma , \theta) =  M_{SS}^2 \left(1- A\, {\rm cos}\,{\tilde \theta }\right)\sigma^2 \ ,
\label{Vst}
\end{equation}
where we define
\beq
M_{SS}^2  \equiv  V_4\mu_7 g_s|\hat G|^2 \  \ \ ,\ \ 
|\hat G|^2 \equiv  Z^{-2}(V_4\mu_7)^{-1} (|G|^2+|S|^2)   \ ,
\label{hatG}
\eeq
and ${\tilde \theta}=\theta-Arg(GS)$.
Note that $M_{SS}$ measures the size of the soft terms and hence the inflaton mass. One can estimate the size 
of these  parameters from purely stringy arguments by assuming an isotropic compactification, $Z\simeq g_s\simeq 1$  and standard GUT values for the gauge couplings \cite{higgsotic}.
One finds  
\beqa
|{\hat G}|&\equiv &[Z^{-1}V_4^{-1/2}\mu_7^{-1/2} G_3]\ \simeq \frac {1}{M_p} \ , \\
M_{SS}^2&\equiv &V_4\mu_7 g_s|\hat G|^2\sim 0.05 M_s^4 |\hat G|^2 \ ,
\label{warpping} 
\eeqa
where $M_s$ is the string scale, assumed to be of order $10^{16}-10^{17}$ GeV. In this way one has  an inflaton mass of order
$M_{SS}\simeq 10^{13}$ GeV, consistent with the cosmological bounds. 

\begin{figure}[h!]
\begin{center}
{\includegraphics[width=0.35\textwidth]{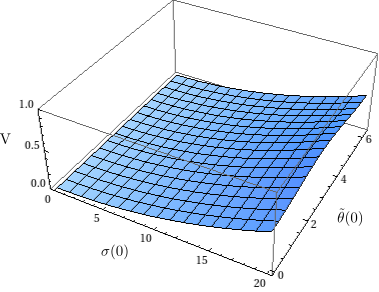}}\quad
{\includegraphics[width=0.35\textwidth]{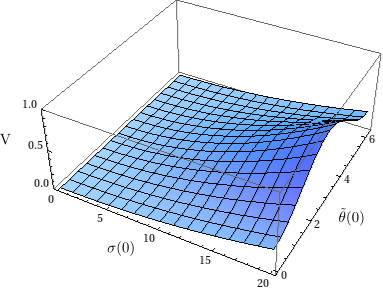}}
{\includegraphics[width=0.35\textwidth]{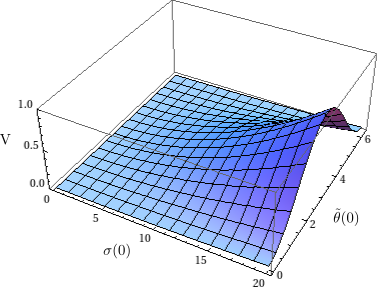}}
\end{center}
\vspace{-15pt}
\caption{Scalar potential for three different values of $A$, $A=0.1$ (upper left), $A=0.5$ (upper right) and $A=0.9$ (below).}
\label{3pot}
\end{figure}

In angular variables the relevant piece of the action may be written as
\beq
\mathcal{L}_{4d}=f(\sigma,\theta)\left(2(\partial_\mu \sigma)^2  +\frac{\sigma^2}{2}(\partial_\mu \theta)^2 \right)- M_{SS}^2(1-A\, {\rm cos}\,\tilde\theta)\sigma^2 \ .
\eeq
with $f(\sigma,\theta)=1\ +\ \frac12 (V_4\mu_7g_s)^{-1}V(\sigma,\theta) $.
Eq.(\ref{Vst}) will be our inflaton potential, although one should not forget that the 
kinetic terms are not canonical and this fact should be taken into account.  This will lead to a general flattening effect, as we discuss below. 
Note that the shape of the scalar potential (before flattening) depends essentially on the value of the $A$-parameter defined above.  In fig.
  \ref{3pot} we show the structure of the scalar potential for three characteristic values $A=0.1, 0.5,0.95$.
%


In the one-field limits, $A=0,1$, one can always get canonical kinetic terms 
by making the following field redefinition

\beq
\phi'=\int d\phi f^{1/2}(\phi).
\label{redef1}
\eeq
In those limits one has a potential
\begin{equation}
V_F\ = \  M_{SS}^2\phi^2\ ,
\end{equation}
with $\phi\equiv\sigma$ for $A=0$ ($G=0$ or $S=0$) or $\phi\equiv |H|$ for $A=1$ ($|G|=|S|$).
The kinetic factor is then
\beq
f\ =\ 1+\frac{|\hat G|^2}{2}\phi^2   \  .
\label{f}
\eeq
From Eq. (\ref{redef1}) one finds the canonically normalised scalar $\varphi$
\beq
\varphi=\frac{1}{2\sqrt{2}} |\phi|\sqrt{2+|\hat G|^2|\phi|^2}+\frac{1}{\sqrt{2}}|\hat G|^{-1}\text{sinh}^{-1}[|\hat G||\phi|/\sqrt{2}]  \ .
\label{redef2}
\eeq
There is not a simple analytic expression for  the inverse function $\phi(\varphi)$,  which is what one would want in order to get the final form of the canonically normalised potential. 
However, for large fields one can check that  the second piece in Eq. (\ref{redef2}) is small and one can invert analytically $\varphi(\phi)$  and
obtain the scalar potential for the new variable
\beq
V_F \  =  \ (g_sV_4\mu_7)|{\hat G}|^2H^2 \ \simeq \  (g_sV_4\mu_7) \left( \sqrt{1\ +\ 8|{\hat G}|^2 \varphi^2}\ -\ 1\right) \ \ .
\eeq
Note that, for large field, the potential becomes linear in the inflaton $\varphi$. This behaviour for large field is also 
present in the results of the complete 2-field analysis below. 
However, finding such an analytic expression is not always possible. In general,
for multiple fields there does not exist a transformation that makes the metric flat on the moduli space parametrized by $\phi^a$.
Therefore, to make a complete analysis we will have to use the general Lagrangian of multiple fields
\beq
\mathcal{L}_{4d}=\frac12 G_{ab}(\phi)\partial_\mu \phi^a \partial^\mu\phi^b- V(\phi) \ , 
\label{def} 
\eeq
where in our case the metric will be  given by
\beq
G_{ab}=\left(\begin{array}{cc}4f(\sigma,\theta)&0\\0&\sigma^2 f(\sigma,\theta)\end{array}\right) \ ,
\label{metric}
\eeq
and use the generalized expressions of the cosmological observables for non-canonical kinetic terms and multiple fields introduced in the next chapter.

\section{Basics of 2-field inflation}\label{sec:2field}
We are interested in studying cosmological perturbations of a system of two scalar fields coupled to Einstein gravity. The action for such a system, if one allows for a curved field space, takes the  form ($8 \pi G = M_{Pl}^2=1$):
\begin{align}
S &=  \int \sqrt{-g} d^4 x \left(\frac{1}{2}R- \frac12 G_{ab}g^{\mu\nu}\partial_{\mu}\phi^a\partial_{\nu}\phi^b-V(\phi)\right) \ ,
\label{eq: infact}
\end{align}
where $g_{\mu\nu}$ is the spacetime metric with determinant $g$ and $G_{ab}$ the field-space metric. The scalar fields $\phi^1$ and $\phi^2$ span a 2-dimensional manifold with all relevant information contained in $G_{ab}$. The Christoffel symbols, Riemann tensor and Riemann scalar of the scalar manifold are defined in the usual way. Assuming that the scalar fields depend only on (cosmic) time, $\phi^a= \phi_0^a(t)$ and that the spacetime metric is the flat FRW metric:
\begin{align}
ds^2  &=  -dt^2+a^2(t) dx^i dx_i \ ,
\end{align} 
the equations of motion for the scalar fields are given by:
\begin{align}
\ddot{\phi}_0^a + \Gamma^a_{bc}\dot{\phi}_0^b\dot{\phi}_0^c + 3 H \dot{\phi}^a_0 + G^{ab}\partial_b V  &=  0 \ .
\label{eq: eombg}
\end{align} 
This can expression can be written more economically using the covariant derivative defined as: $D\dot{\phi}^a=d\dot{\phi}_0^a+\Gamma^a_{bc}\dot{\phi}_0^b\dot{\phi}_0^c$. The function $a$ is the scale factor and the Hubble parameter, $H$, is defined as: $H=\partial_t \text{ln}(a)$. The equation of motion for the scale factor is:
\begin{align}
H^2  &=  \frac13 \left( \frac12 \dot{\phi}_0^2+V\right)\ ,
\label{eq: eoma}
\end{align}
where $\dot{\phi}_0^2 = G_{ab}\dot{\phi}_0^a\dot{\phi}_0^b$. Solutions to Eqs. (\ref{eq: eombg}) and (\ref{eq: eoma}) determine the background evolution of the system.

It is useful to define the slow-roll parameters:
\begin{align}
\epsilon  &\equiv  - \frac{\dot{H}}{H^2}\ , \\
\eta  &\equiv  \frac{\dot{\epsilon}}{H\epsilon}\ .
\end{align}
Physically, $\epsilon$ measures the change of the Hubble sphere of the local universe. Hence, $0<\epsilon<1$ corresponds to a shrinking Hubble sphere, which is a good definition of an inflationairy period. In order for inflation to last a sufficient amount of time the derivative of $\epsilon$ has to be small as well, so $\eta\ll1$. It is possible to use the equations of motion (\ref{eq: eombg}) and (\ref{eq: eoma}) to relate $\epsilon$ to $\dot{\phi}^2_0\equiv G_{ab}\dot{\phi}_0^a\dot{\phi}_0^b$, which implies that $\eta$ is related to the tangential acceleration of the background trajectory.

In order to make connection with observations it is necessary to consider quantum perturbations around classical background solutions. This involves expanding the scalar and metric degrees of freedom in terms of the background quantities ($\phi_0^a$ and $g^0_{ab}$) and perturbations, finding the equations of motion for the gauge invariant perturbations and solving them \cite{Sasaki:1995aw}. The spectral index and tensor-to-scalar ratio are defined in terms of the power spectra of the quantum fluctuations of scalar and tensor modes as we will shortly review.

It is convenient to define a local frame on the trajectory in field space given by \cite{GrootNibbelink:2000vx,GrootNibbelink:2001qt, Achucarro:2010da}:
\begin{align}
T \quad &= \quad \frac{1}{\dot{\phi}_0} (\dot{\phi}_0^1,\quad \dot{\phi}_0^2)\ , \\
N \quad &= \quad \frac{1}{\sqrt{G}\dot{\phi}_0}(-G_{22}\dot{\phi}_0^2-G_{12}\dot{\phi}_0^1,\quad G_{11}\dot{\phi}_0^1+G_{12}\dot{\phi}_0^2)\ .
\label{eq: basis}
\end{align}
These vectors form an orthonormal basis of the tangent space of the scalar manifold. Hence, they can be used to decompose the physically relevant quantities along the normal and tangential directions with respect to the background trajectory. In particular, the derivatives of the scalar potential can be written in this basis as  $V_{\phi}=T^a \partial_a V$ and $V_N=N^a\partial_a V$ ($a$ labels both components of the basis vectors as well as fields). The  total acceleration of the fields can be derived by taking the covariant derivative of $T^a$, using the equations of motion and projecting with $N^a$. One obtains:
\begin{align}
\ddot{\phi}_0 \quad &= \quad -3H \dot{\phi}_0 - V_{\phi}\ .
\end{align}
The parameter $\eta$ needs to be generalised  to capture the full dynamics in field space. Recalling that in single field settings $\eta$ measures the acceleration of the  scalar field, one finds that its convenient multifield generalisation is
\begin{align}
\eta^a \quad &= \quad -\frac{1}{H \dot{\phi}_0} D \dot{\phi}_0^a\ ,
\end{align} 
which measures the acceleration for the field $\phi^a$.

In the local $(T,N)$ basis, $\eta^a$ is projected on the following parallel and perpendicular components:
\begin{align}
\eta_{\parallel} \quad &= \quad - \frac{\ddot{\phi}_0}{H\dot{\phi}_0}\ , \\
\eta_{\perp} \quad &= \quad \frac{V_N}{H\dot{\phi}_0}\ ,
\end{align}
such that:
\begin{align}
\eta^a \quad &= \quad \eta_{\parallel}T^a+ \eta_{\perp}N^a\ .
\end{align}
$\eta_\parallel$ measures the tangential acceleration, responsible for the variation of the modulus of the background trajectory velocity whereas $\eta_\perp$ measures the normal acceleration causing the background trajectory to curve.

The equations of motion for the perturbations are given in terms of the gauge-invariant Mukhanov-Sasaki variables \cite{Sasaki:1986hm,Mukhanov:1988jd}:
\begin{align}
Q^a \quad &= \quad \delta \phi^a + \frac{\dot{\phi}^a}{H}\psi\ ,
\end{align}
which can also be decomposed using the basis of Eq. (\ref{eq: basis}) as :
\begin{align}
v^T \quad &= \quad a T_a Q^a\ , \\
v^N \quad &= \quad a N_a Q^a\ .
\end{align}
Quantisation of these variables is  done by the standard procedure, defining the canonical momenta and demanding the correct commutation relations. The variables are given in Fourier space by:
\begin{align}
v^{N,T}(x,\tau) \quad &= \quad \int \frac{d^3k}{(2 \pi)^{3/2}} e^{ik \cdot x} \sum_{\alpha} \left( v^{N,T}_{\alpha}(k,\tau) a_{\alpha}(k)+v_{\alpha}^{N,T *}(k,\tau)a^{\dagger}_{\alpha}(-k)\right),
\end{align}
where $a_{\alpha}$ and $a_{\alpha}^{\dagger}$ are the usual creation and annihilation operators. We have switched to conformal time defined by: $d\tau=1/a(t) dt$. The Greek index $\alpha$ labels the quantum modes of the perturbations, consistency requires that $\alpha=1,2$ in the 2-field case. 

The equations of motion for the Mukhanov-Sasaki variables in this basis take the form \cite{Achucarro:2010da}:
\begin{align}
\frac{d^2 v^T_{\alpha}}{d\tau^2} + 2 a H \eta_{\perp} \frac{dv_{\alpha}^N}{d\tau}-a^2 H^2 \eta^2_{\perp}v_{\alpha}^T+\frac{d(aH\eta_{\perp})}{d\tau}v_{\alpha}^N+\Omega_{TN}v_{\alpha}^N+(\Omega_{TT}+k^2)v_{\alpha}^T &= 0, \label{eq:MStau1}\\
\frac{d^2 v^N_{\alpha}}{d\tau^2} - 2 a H \eta_{\perp} \frac{dv_{\alpha}^T}{d\tau}-a^2 H^2 \eta^2_{\perp}v_{\alpha}^N-\frac{d(aH\eta_{\perp})}{d\tau}v_{\alpha}^T+\Omega_{TN}v_{\alpha}^T+(\Omega_{NN}+k^2)v_{\alpha}^N &= 0. 
\label{eq:MStau2}
\end{align}
We see that the 2-field perturbation system consists of a set of pairwise coupled harmonic oscillators. The coupling between curvature ($ v^T_{\alpha}$) and isocurvature ($ v^N_{\alpha}$) modes is controlled by $\eta_\perp$ which is inversely proportional to the curvature radius of the background trajectory. It follows that the coupling between curvature and isocurvature will be strong whenever there is a sharp turn in the background trajectory.  The symmetric mass matrix $\Omega$ of Eq. \eqref{eq:MStau1} and \eqref{eq:MStau2} has the following elements:
\begin{align}
\Omega_{TT}\quad &= \quad -a^2H^2 (2+2\epsilon-3\eta_{\parallel}+\eta_{\parallel}\xi_{\parallel}-4\epsilon\eta_{\parallel}+2\epsilon^2-\eta^2_{\perp})\ , \\
\Omega_{NN}\quad &= \quad -a^2H^2(2-\epsilon)+a^2 V_{NN}+a^2H^2\epsilon R\ , \\
\Omega_{TN} \quad &= \quad a^2 H^2 \eta_{\perp}(3+\epsilon-2\eta_{\parallel}-\xi_{\perp})\ ,
\end{align}
where R is the Ricci scalar of the scalar manifold and the third slow-roll parameters are defined as:
\begin{align}
\xi_{\parallel} \quad &= \quad - \frac{\dddot{\phi}_0}{H \ddot{\phi}_0}\ , \\
\xi_{\perp} \quad &= \quad -\frac{\dot{\eta}_{\perp}}{H \eta_{\perp}} \ .
\end{align}

Assuming that the system is decoupled  ($\eta_\perp=0$)  when observationally relevant modes are deep inside the horizon ($\frac{k}{a H}\gg 1$), canonical quantisation  fixes the initial conditions for the scalar perturbations to be
\begin{align}
v_{\alpha}^{N,T}\quad &= \quad \delta_{\alpha}^{N,T} \frac{1}{\sqrt{2k}}e^{-ik\tau}\ , 
\label{eq:BD}
\end{align}
i.e. the initial conditions are simply given by the Bunch-Davies vacuum. Note that in Eq. \eqref{eq:BD} it is understood that $\delta^T_1=1$ etc. and derivatives define the initial condition for $\frac{d}{d\tau}v_{\alpha}^{T,N}$. It is important to note that there are two sets of equations of motion for the Muhkanov-Sasaki variables (in total 4 equations) and corresponding initial conditions, one for each value of $\alpha$ and that both should be taken into account when computing the  inflationary observables. 

The power spectra are defined in terms of the scalar 2-point functions as 
\begin{align}
P_{\zeta}(k,\tau) \quad &= \quad \frac{k^3}{4 \pi^2 a^2 \epsilon} \sum_{\alpha=1,2} v_{\alpha}^T(k,\tau)v_{\alpha}^{T*}(k,\tau) \ , \label{eq:Pzeta}\\
P_{\mathcal{S}}(k,\tau) \quad &= \quad \frac{k^3}{4 \pi^2 a^2 \epsilon} \sum_{\alpha=1,2} v_{\alpha}^N(k,\tau)v_{\alpha}^{N*}(k,\tau) \ .
\label{eq:PS}
\end{align}
where $P_{\zeta}(k,\tau)$ and $P_{S}(k,\tau)$ denote the dimensionless power spectra for the curvature and isocurvature modes respectively. Given that in multi-field models there can be superhorizon evolution of the perturbations, these are to be evaluated at the end of inflation. This is to be contrasted to single field models, where the freezing of curvature perturbations on superhorizon scales means the power spectra can be evaluated at horizon exit. From Eq. \eqref{eq:Pzeta} one can compute the spectral index for the curvature perturbations
\begin{align}
n_s \quad &= \quad 1+ \frac{d\text{ ln}(P_{\zeta}(k,\tau_{end}))}{d \text{ ln}(k)} \ ,
\end{align}
as well as the amplitude at the pivot scale $k_*$
\be
A_s=P_{\zeta}(k_*,\tau_{end}),
\ee
which in the absence of an analytical solution to Eqs. \eqref{eq:MStau1}, \eqref{eq:MStau2} must be computed numerically. Noting that the tensor modes' evolution is unaffected by the number of dynamical fields driving the background expansion, the amplitude of the tensor power spectrum is given as in the single field case by
\be
A_t=P_{t}(k_*,\tau_{end})=\frac{2 }{\pi}H^2 \ ,
\ee
which implies the following definition of the tensor to scalar ratio
\be
r\equiv\frac{P_{t}}{P_\zeta}(k_*,\tau_{end})=\frac{2 H^2}{\pi P_{\zeta}(k_*,\tau_{end})} \ .
\ee
Besides probing the scalar and tensor power spectra, observations also put bounds on the total fraction of primordial isocurvature, defined as 
\be
\beta_{iso}=\frac{P_{\mathcal{S}}}{P_{\mathcal{S}}+P_\zeta} \ .
\ee
From the theoretical point of view the isocurvature fraction depends on the mass of the isocurvature modes ($\Omega_{NN}$) and on the strength of their coupling to the adiabatic perturbations. The observational bounds on $\beta_{iso}$ can vary by many orders of magnitude, depending on how primordial isocurvature is transferred to the post-inflationary Universe. From \cite{Ade:2015lrj} we find that the less constraining bound is of the order
\be
\beta_{iso}\leq 10^{-3}
\ee
at the end of inflation. 

In addition to putting constraints on the fraction of primordial isocurvature, observations also put constraints on the non-linear non-Gaussianity parameters $f_{NL}$. Producing large non-Gaussianities would, in principle, spoil the validility of the model.  However, for 2-field models the non-linear $f_{NL}$ are of the order of the slow-roll parameters and hence they are surpressed in our model, see \cite{Vernizzi:2006ve}, \cite{Yokoyama:2007uu}. We will not consider non-Gaussianities beyond this point.

\subsection{Decoupling limit and the single field observables}
In order to understand how to relate the observables defined above with those of single field inflation let us once more take the decoupling limit $\eta_{\perp}\rightarrow 0$ and reduce the  equations of motion to 
\bea
\frac{d^2 v^T_{\alpha}}{d\tau^2}+\left[ k^2+\frac{1}{\tau^2}(-2-6\epsilon+3\eta_\parallel) \right]v_{\alpha}^T &= 0 \ , 
\label{eq:MStauDec1}\\
\frac{d^2 v^N_{\alpha}}{d\tau^2} +\left[k^2 +\frac{1}{\tau^2}\left(-2+\frac{M^2}{H^2}+\left(-3+\frac{2 M^2}{H^2}\right) \epsilon \right)\right]v_{\alpha}^N &= 0 \ ,
\label{eq:MStauDec2}
\eea
where the isocurvature mass is $M^2=V_{NN}+H^2 \epsilon R$ and we have used $\tau^{-1}=a H (1-\epsilon)$ for the background evolution. Equations \eqref{eq:MStauDec1} and \eqref{eq:MStauDec2} admit solutions that are a superpositions of Hankel functions of the first and second kind, which upon imposing Bunch-Davies boundary conditions reduce to
\begin{align}
 v^T_{1}=& \frac{\sqrt{-\tau \pi}}{2}\: \: H^{(1)}_{\nu_T}(-k \tau)\equiv v^T\qquad,\qquad \nu_T=\frac{3}{2}+2\epsilon-\eta_\parallel\\
 v^N_{2}=&\frac{\sqrt{-\tau \pi}}{2}\: \: H^{(1)}_{\nu_N}(-k \tau)\equiv v^N\qquad,\qquad \nu_N=\frac{3}{2}\sqrt{1 -\left(\frac{2 M}{3 H}\right)^2 (1+2\epsilon)+\frac{4}{3}\epsilon}
\label{eq:MSsols}
\end{align}
up to overall unimportant phases, with $ v^T_{2}= v^N_{1}=0$. Note that $\nu_T=3/2$ and $\nu_N<3/2$ to zeroth order in the slow-roll expansion.

On superhorizon scales, in the decoupling limit, one can show that the curvature perturbations are frozen as in the pure single field case, since $Q^T=v^T/a \propto a^{0}$ while the isocurvature perturbations decay as $Q^N=v^N/a\propto a^{\nu_N-3/2}\sim a^{-\frac{2 M^2}{9 H^2}}$ to zeroth order in slow roll.
 
In order to make contact with observations one can compute the dimensionless power spectrum of curvature perturbations in this limit, finding
\be
P_\zeta=\frac{k^3}{4\pi^2 a^2 \epsilon}|v_T|^2\rightarrow \left(\frac{k}{aH}\right)^{3-2 \nu_T}\frac{H^2}{8 \pi \epsilon}\ ,
\ee
on superhorizon scales, which implies the following definitions for the spectral index and the amplitude 
\be
n_s-1=3-2\: \nu_T=-4\:\epsilon+2\: \eta_\parallel=-2 \epsilon-\eta \qquad\text{and}\qquad A_s=\frac{H^2}{8 \pi^2 \epsilon} \ .
\label{eq:nsSingle}
\ee
Given that in the $\eta_\perp=0$ limit the curvature perturbations are frozen once they leave the horizon, these observables can be evaluated at horizon exit, when $k=aH$.

By performing a similar computation for the isocurvature modes one can show that the amplitude of the isocurvature power spectrum at horizon crossing is the same as for the adiabatic modes. As noted above, due to the fact that $\nu_N<3/2$, the isocurvature perturbations decay on superhorizon scales with a rate controlled by the ratio $M^2/H^2$. This implies that the isocurvature fraction at the end of inflation scales as
\be
\beta_{iso}\sim \frac{|v^N|^2}{|v^T|^2}=a^{-\frac{4 M^2}{9 H^2}}
\ee
and is therefore suppressed if at some stage during observable inflation $M \ge H$.

The tensor to scalar ratio is given in the decoupling limit by its single field expression
\be
r=\frac{A_t}{A_s}=16\: \epsilon_* \ ,
\label{eq:rSingle}
\ee
where $*$ denotes evaluation at horizon crossing since both adiabatic and tensor perturbations are frozen outside the horizon.

\section{Results}\label{sec:results}

Having reviewed the formalism for the computation of inflationary observables in 2-field models we now reanalyse Higgs-otic inflation in three representative points in parameter space: $A=0.83$, $A=0.7$ and $A=0.2$. First, we will keep the total amount of flux fixed at $\hat{G}=1$. For standard values of the string scale this leads to a supersymmetry breaking scale of $M_{SS}\simeq 10^{13}$ GeV.
However, we will also show the results for a large range of ${\hat G}$ values at the end of this chapter.


\subsection{Higgs-otic  regime: $A=0.83$, $\hat{G}=1$} \label{sec:A083}

In this section, we present the results for the inflationary observables corresponding to the {\it canonical}  flux choice $A=0.83$ and $\hat{G}=1$. For this point in parameter space, at the   SUSY-breaking scale $M_{SS}\simeq 10^{13}$ GeV, there exist a heavy Higgs that can be integrated out and a light (approximately massless) Higgs, to be identified with the Standard Model Higgs  field. As mentioned in Section \ref{sec:Higgsotic}, at the ultraviolet (string)  scale the two fields have comparable masses (in fact for $A=0.83$ one has $m_H/m_h=3.28$), implying that inflationary dynamics driven by such a Higgs sector will necessarily be multi-field in nature and that the observables are better estimated via the methods reviewed in Section \ref{sec:2field}.

\begin{figure}[h!]
	\centering
	\begin{minipage}[b]{0.49\linewidth}
	\centering
	\includegraphics[width=1\textwidth]{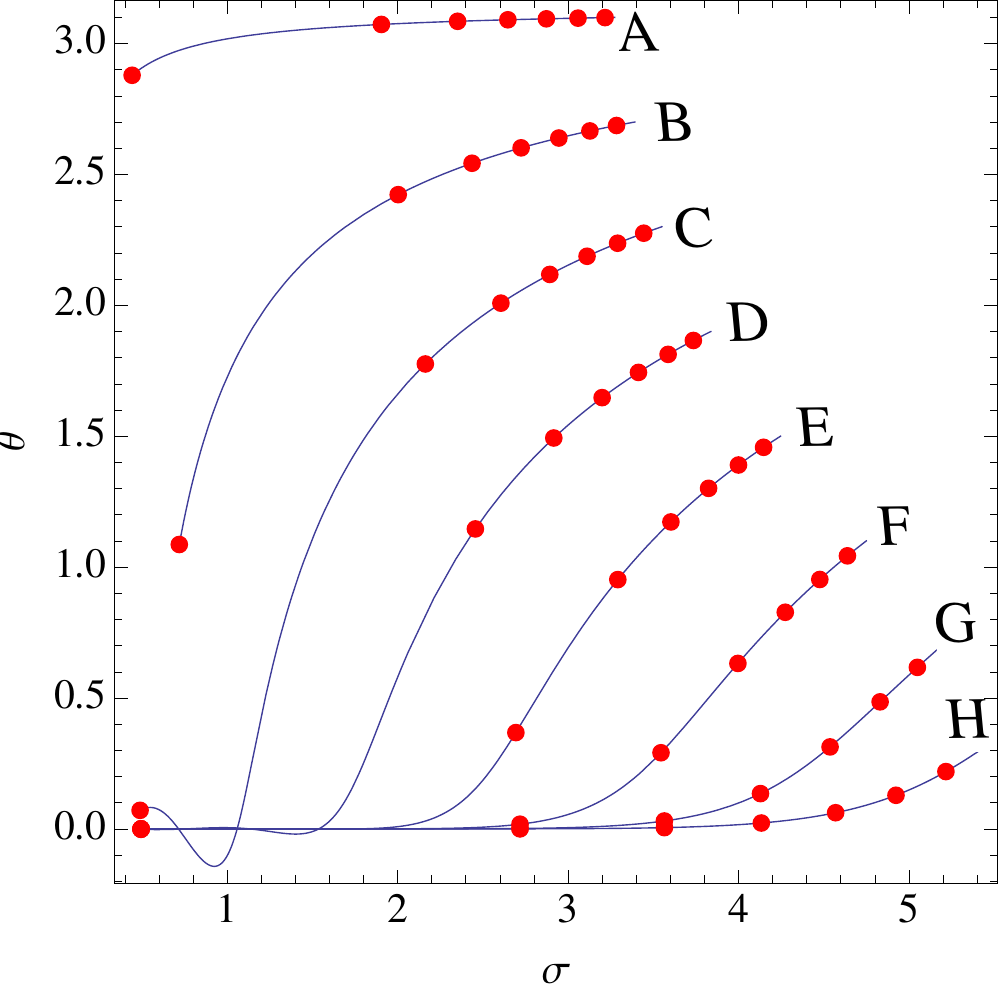}
    \end{minipage}
	\hspace{0.05cm}
	\begin{minipage}[b]{0.49\linewidth}
	\centering
\includegraphics[width=1\textwidth]{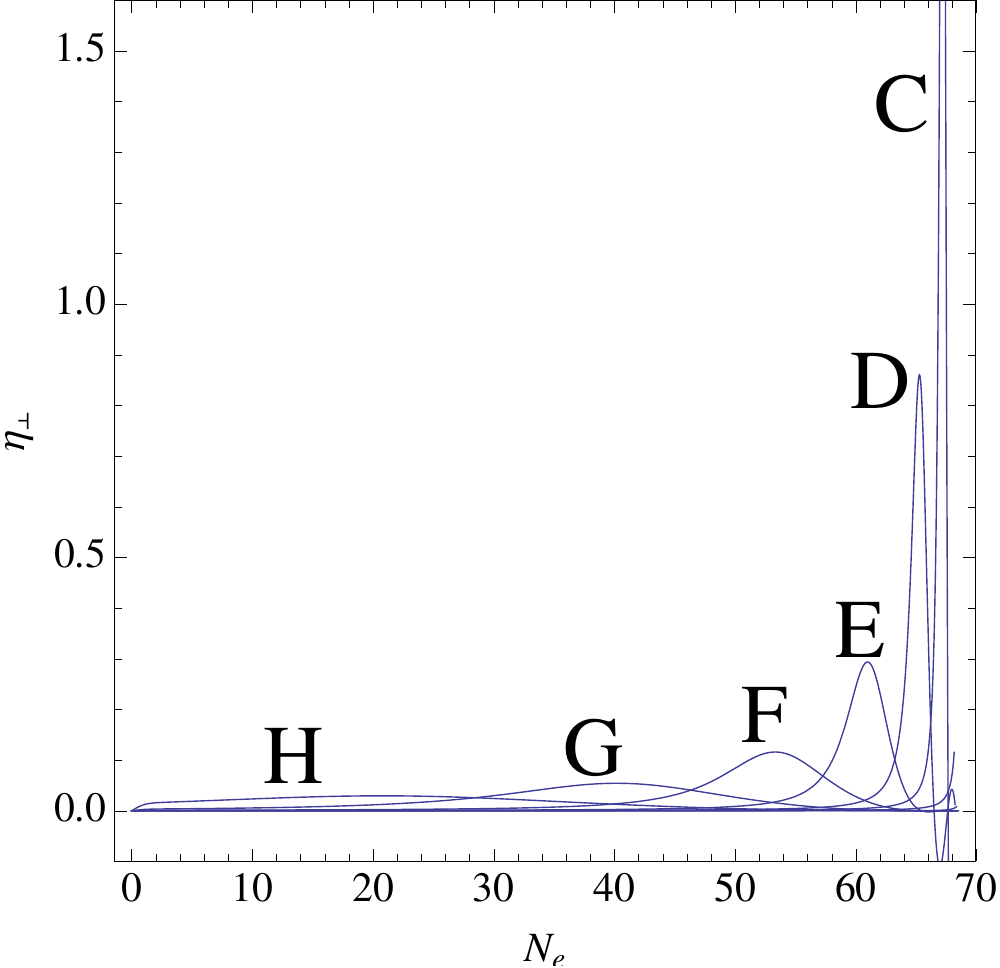}
	\end{minipage}
	\hspace{0.05cm}
\caption{Left: The last 68 efoldings for some representative inflationary trajectories. Red dots mark 10 efoldings intervals on each trajectory. Right: Evolution of $\eta_\perp$ for the different trajectories.}
	\label{fig:BGGhat1}
\end{figure}

We present in figure \ref{fig:BGGhat1} several trajectories in the $(\sigma,\theta)$ plane as well as the evolution of the all important $\eta_\perp$ parameter for those same trajectories. Recalling that $\eta_\perp$ is proportional to the inverse curvature radius of the background trajectory \cite{Achucarro:2010da}, we observe that the marked turns in the trajectories generate clear peaks in $\eta_\perp$. These peaks are sharper and higher for trajectories where the turning takes place close to the end of inflation (large $\theta_0$ trajectories).

The multi-field effects, arising through the $\eta_\perp$ controlled coupling between curvature and isocurvature modes, are more severe the earlier the turn takes place. So even though the coupling is stronger for say trajectory C than for trajectory H, its effects are more pronounced in the latter since the isocurvature and curvature modes are coupled at earlier times, when there is more isocurvature power. 

\begin{figure}[h!]
	\centering
	\begin{minipage}[b]{0.49\linewidth}
	\centering
	\includegraphics[width=0.95\textwidth]{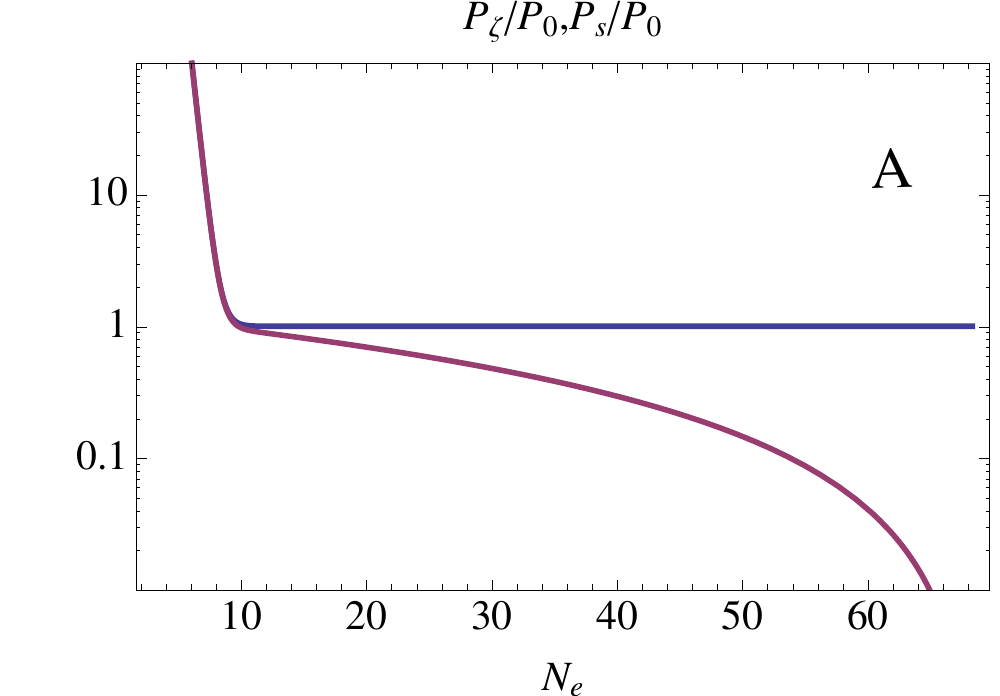}
    \end{minipage}
	\hspace{0.05cm}
	\begin{minipage}[b]{0.49\linewidth}
	\centering
\includegraphics[width=0.95\textwidth]{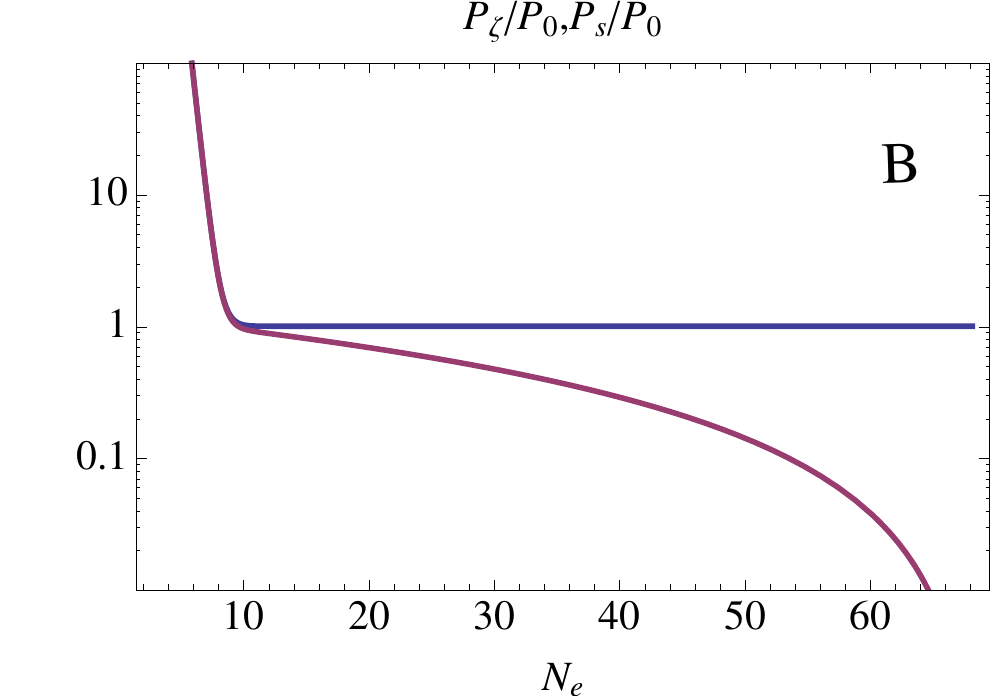}
	\end{minipage}
	\hspace{0.05cm}
\centering
	\begin{minipage}[b]{0.49\linewidth}
	\centering
	\includegraphics[width=0.95\textwidth]{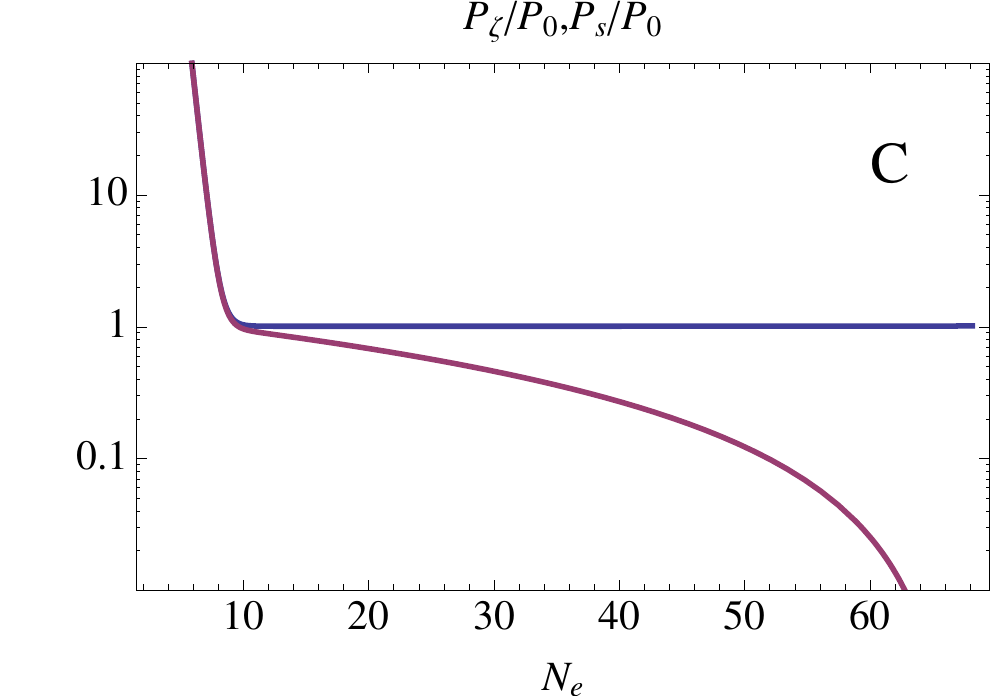}
    \end{minipage}
	\hspace{0.05cm}
	\begin{minipage}[b]{0.49\linewidth}
	\centering
\includegraphics[width=0.95\textwidth]{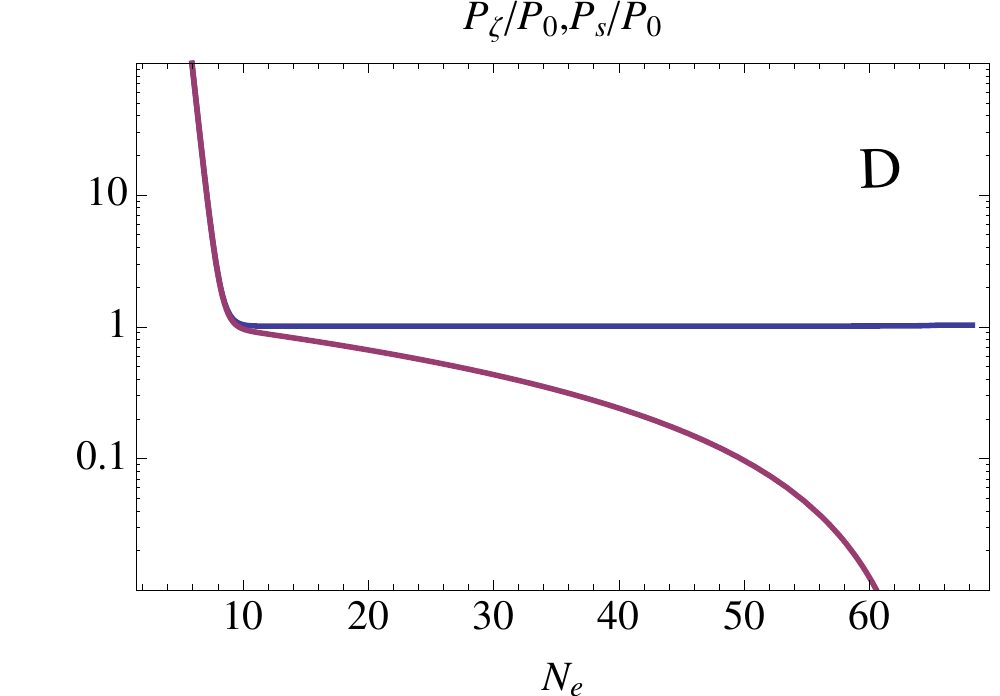}
	\end{minipage}
	\hspace{0.05cm}
\centering
	\begin{minipage}[b]{0.49\linewidth}
	\centering
	\includegraphics[width=0.95\textwidth]{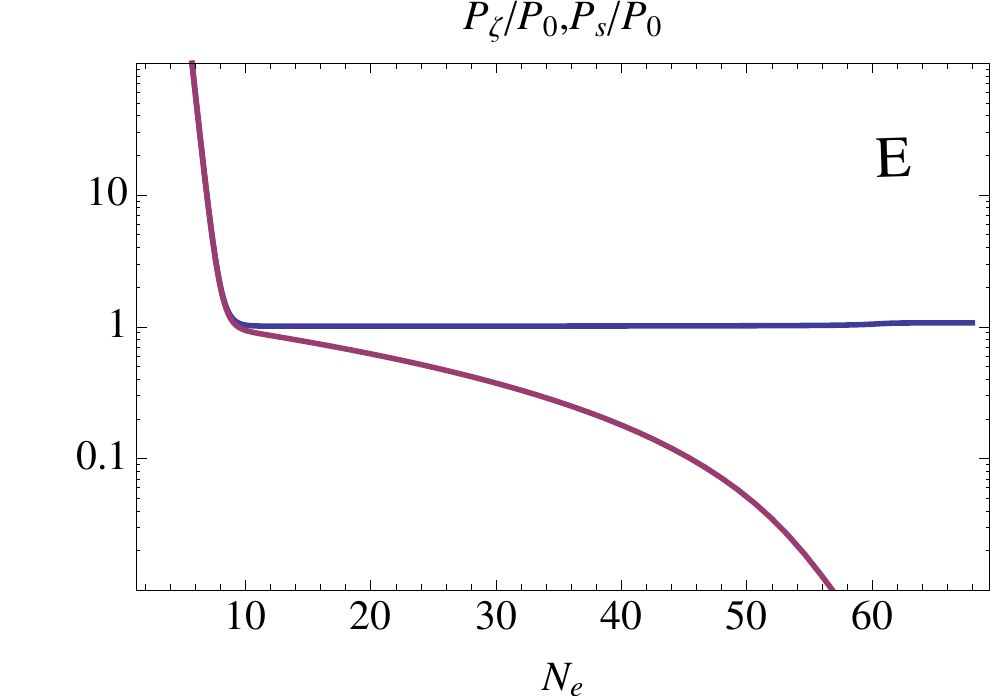}
    \end{minipage}
	\hspace{0.05cm}
	\begin{minipage}[b]{0.49\linewidth}
	\centering
\includegraphics[width=0.95\textwidth]{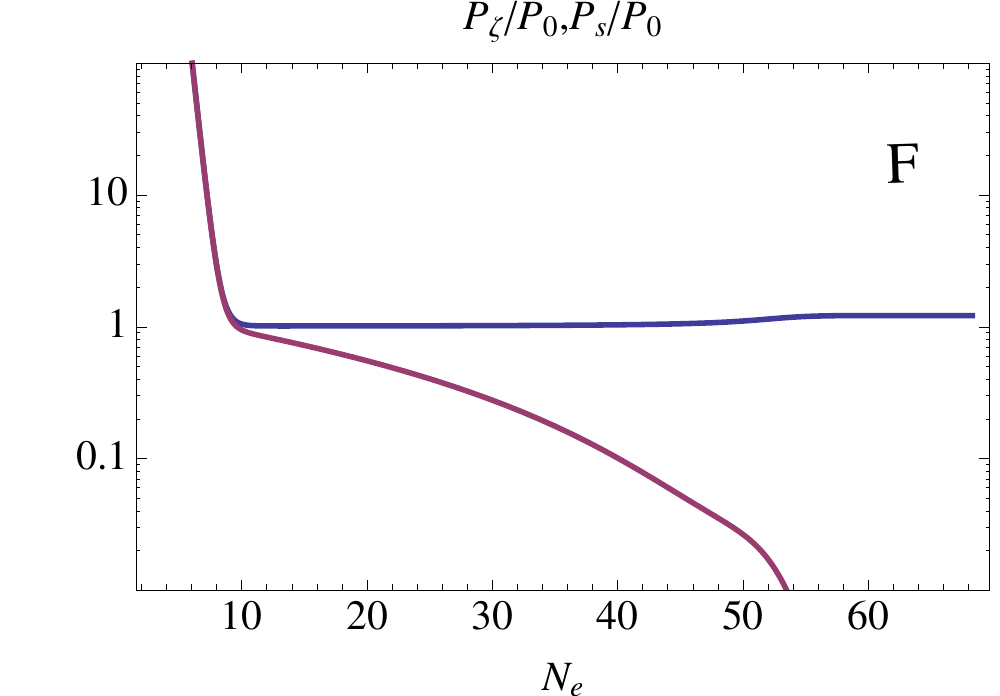}
	\end{minipage}
	\hspace{0.05cm}	
\centering
	\begin{minipage}[b]{0.49\linewidth}
	\centering
	\includegraphics[width=0.95\textwidth]{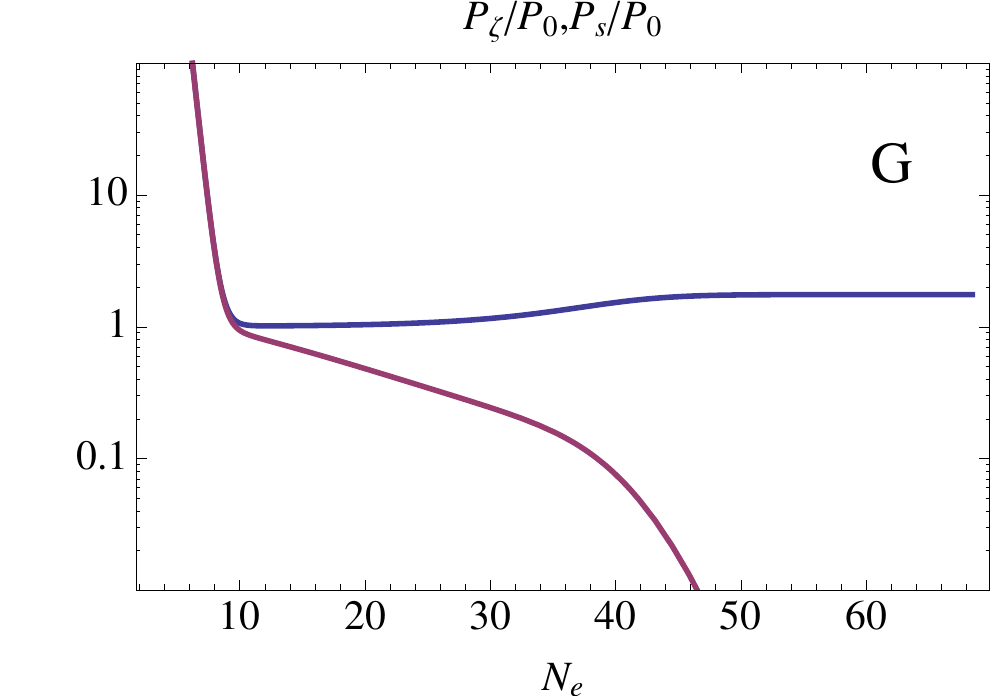}
    \end{minipage}
	\hspace{0.05cm}
	\begin{minipage}[b]{0.49\linewidth}
	\centering
\includegraphics[width=0.95\textwidth]{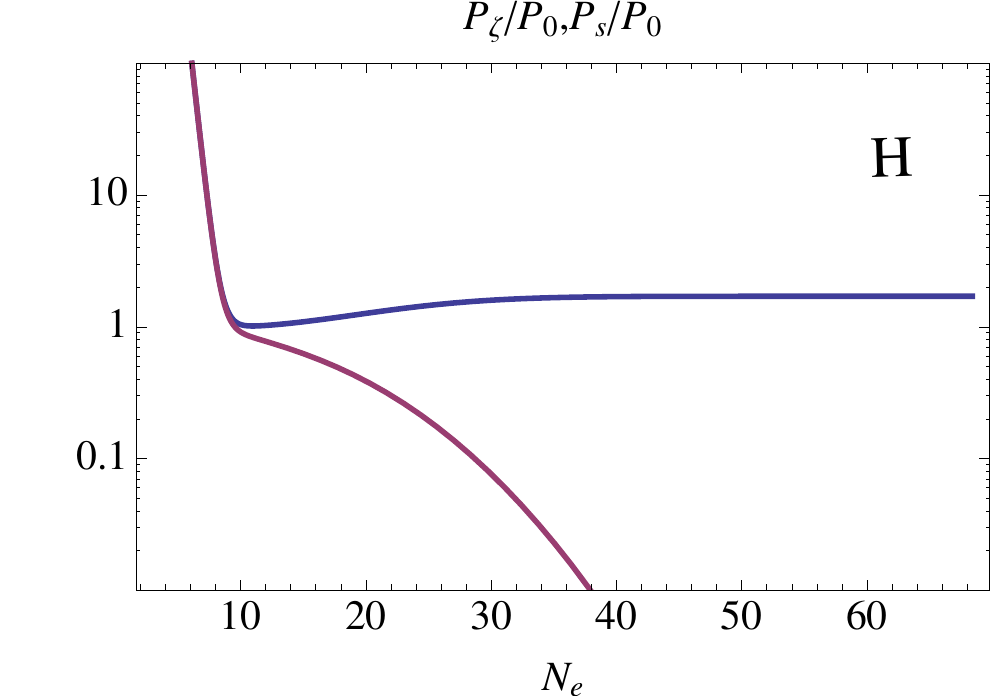}
	\end{minipage}
	\hspace{0.05cm}	
\caption{Evolution of the curvature (blue) and isocurvature (red) two point functions for $k_{60}$ for the trajectories of Fig. \ref{fig:BGGhat1}. Curvature and isocurvature power are normalised to the single field estimate $P_0=\frac{H^2}{8 \pi^2 \epsilon}\big|_{*}$.}
	\label{fig:PPSGhat1}
\end{figure}

The effects of the multi-field dynamics on the amplitude of the scalar curvature perturbations are illustrated in figure \ref{fig:PPSGhat1} where we see that the power transfer from isocurvature to curvature is maximized the earlier the turn takes place. The impact of multi-field effects on the scalar amplitude for trajectories A-E is minimal, with the single field estimate of Eq. (\ref{eq:nsSingle}) providing a good approximation to the full result. For trajectories F-H the superhorizon evolution of the curvature perturbations driven by isocurvature power transfer implies that Eq. (\ref{eq:nsSingle})  underestimates the amplitude by as much as $80\%$. These results are summarised in table \ref{tab:betaIsoGhat03}, where we also present the estimates for the primordial isocurvature fraction, $\beta_{iso}$,  at the end of inflation on the $k_{50}$ and $k_{60}$ scales. We observe that $\beta_{iso}$ varies by many orders of magnitude, being larger for late turning trajectories (large $\theta_0$) where power transfer between isocurvature and curvature perturbations is less efficient and the attenuation of isocurvature power is mostly driven by its decay on superhorizon scales.

\begin{table}[h]
\begin{center}
\begin{tabular}{c|c|c|c|c|c|c|c|c}
\hline
\hline
Trajectory &A&B&C&D&E&F&G&H \\
 \hline
  \hline
 $P_\zeta/P_0\big |_{k_{60}} $&1.01&  1.01&  1.02&  1.03&  1.07&  1.22& 1.76&  1.716\\
 \hline
 $\log_{10} \beta_{iso}(k_{60})$ & -3& -3& -5& -8& -13& -16& -18& -20\\
 \hline
 \hline
 $P_\zeta/P_0\big |_{k_{50}} $ &1.01 & 1.01& 1.02&  1.04 &  1.09&  1.30&  2.00&1.33  \\
\hline
 $\log_{10} \beta_{iso}(k_{50})$& -3 & -3& -5& -8& -13& -16& -18& -19\\
  \hline 
  \hline
\end{tabular}
\end{center}
\label{tab:betaIsoGhat03}
\caption{Ratio between the amplitude of the curvature perturbations at the end of inflation and the single field estimate of eq. \eqref{eq:nsSingle} for the trajectories of Fig. \ref{fig:BGGhat1}.}
\end{table}%

\begin{figure}[h]
	\centering
	\begin{minipage}[b]{0.49\linewidth}
	\centering
	\includegraphics[width=1\textwidth]{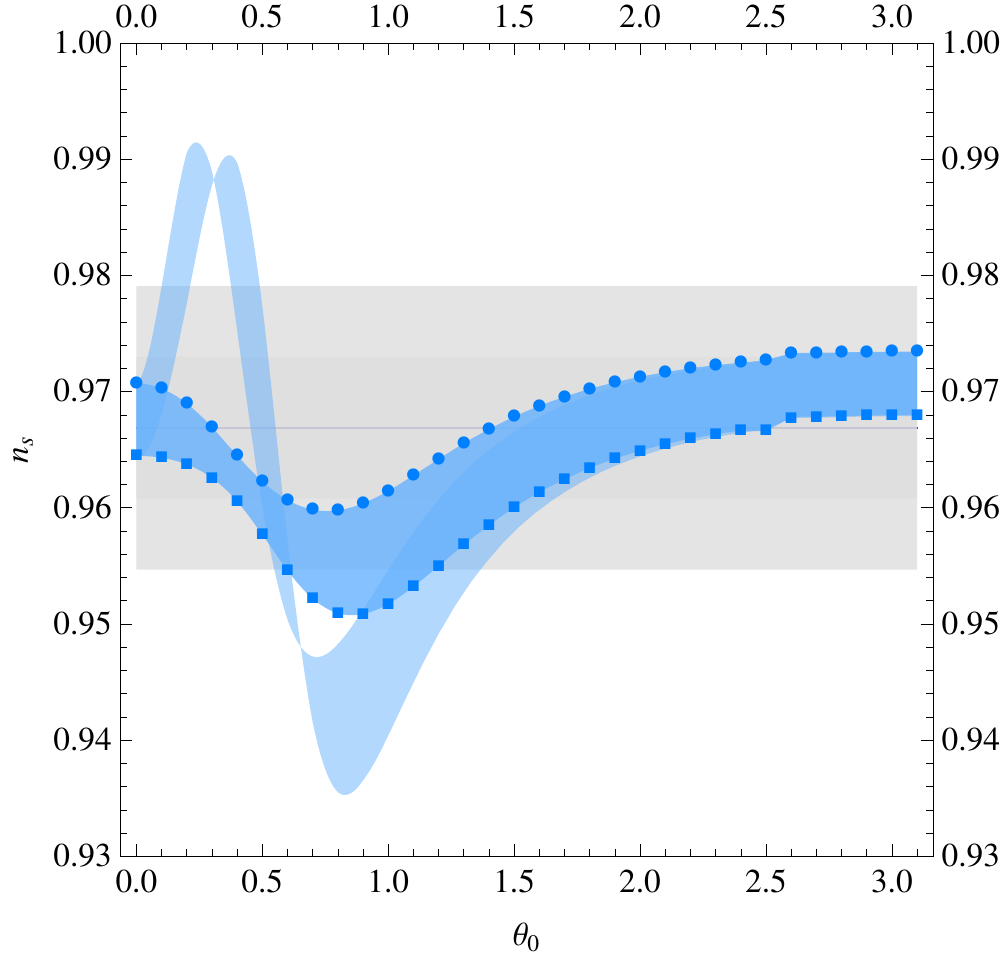}
    \end{minipage}
	\hspace{0.05cm}
	\begin{minipage}[b]{0.49\linewidth}
	\centering
\includegraphics[width=1\textwidth]{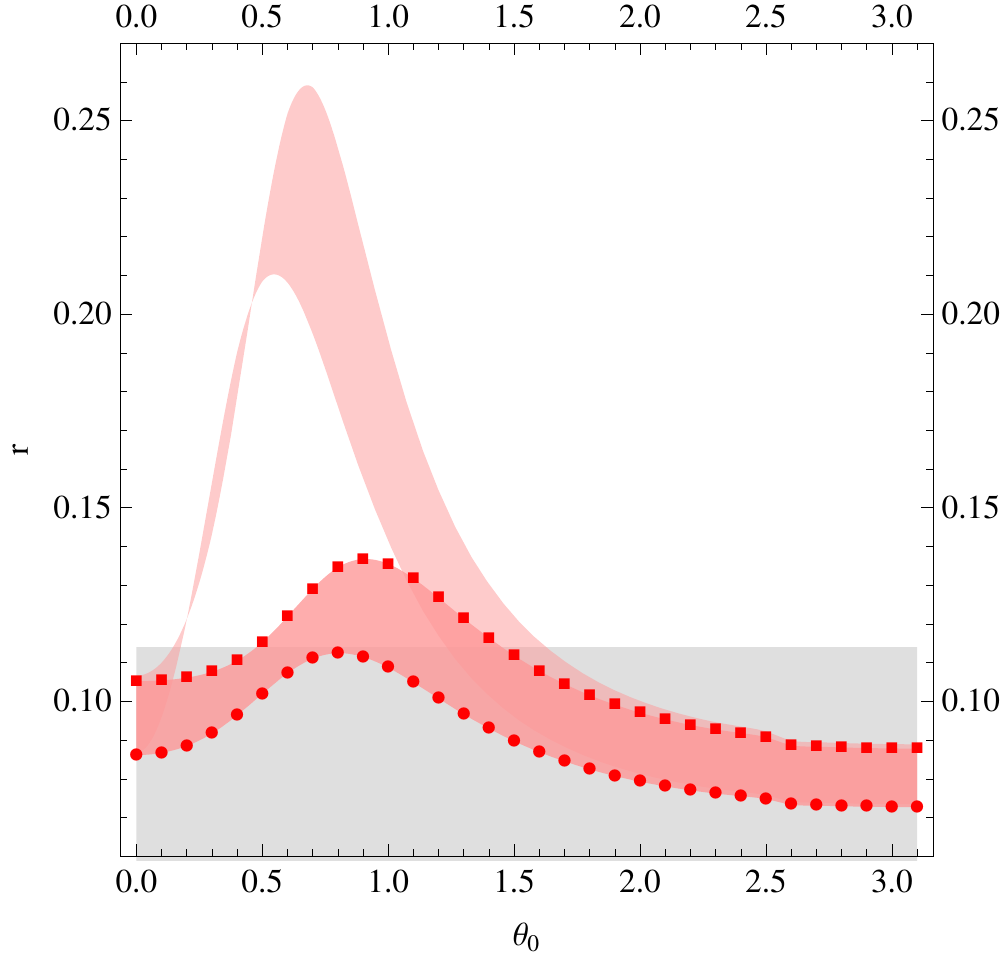}
	\end{minipage}
	\hspace{0.05cm}
\caption{Spectral index (left) and tensor to scalar ratio (right) for $\hat{G}=1$ and $A=0.83$ superimposed with the PLANCK 2015 constraints $n_s= 0.96688 \pm 0.0061$ and $r<0.114$ (grey band). The slightly transparent curve corresponds to the single field estimate while the one surrounded by the data points corresponds to the multi-field results. Circles: $N_e=60$, Squares: $N_e=50$.}
	\label{fig:nsrA083}
\end{figure}

In the same way as the multi-field effects can lead to an underestimate of the scalar amplitude, they will also impact other inflationary observables, in particular the tightly constrained spectral index and the tensor to scalar ratio. The single and multi-field estimates for these quantities are plotted as functions of the initial condition $\theta_0$ in figure \ref{fig:nsrA083}. 
\begin{figure}[h]
	\centering
	\begin{minipage}[b]{0.49\linewidth}
	\centering
	\includegraphics[width=1\textwidth]{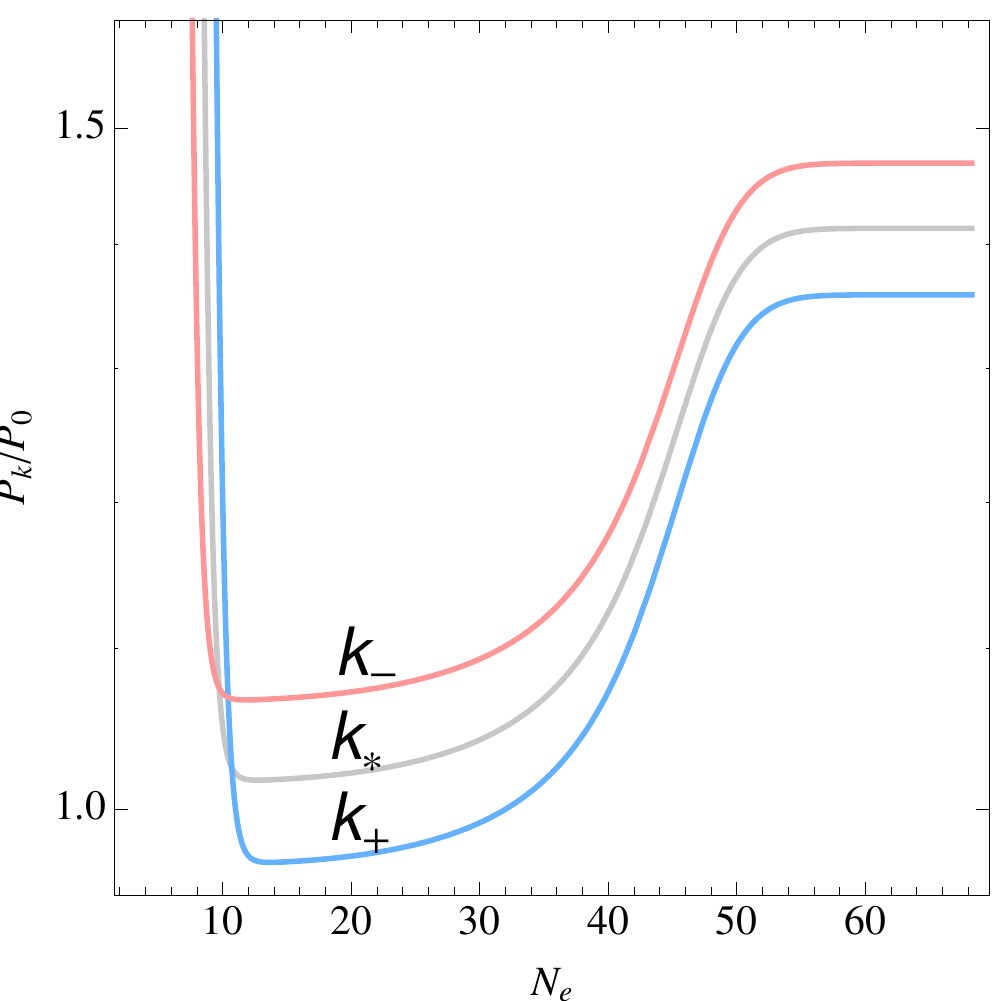}
    \end{minipage}
	\hspace{0.05cm}
	\begin{minipage}[b]{0.49\linewidth}
	\centering
\includegraphics[width=1\textwidth]{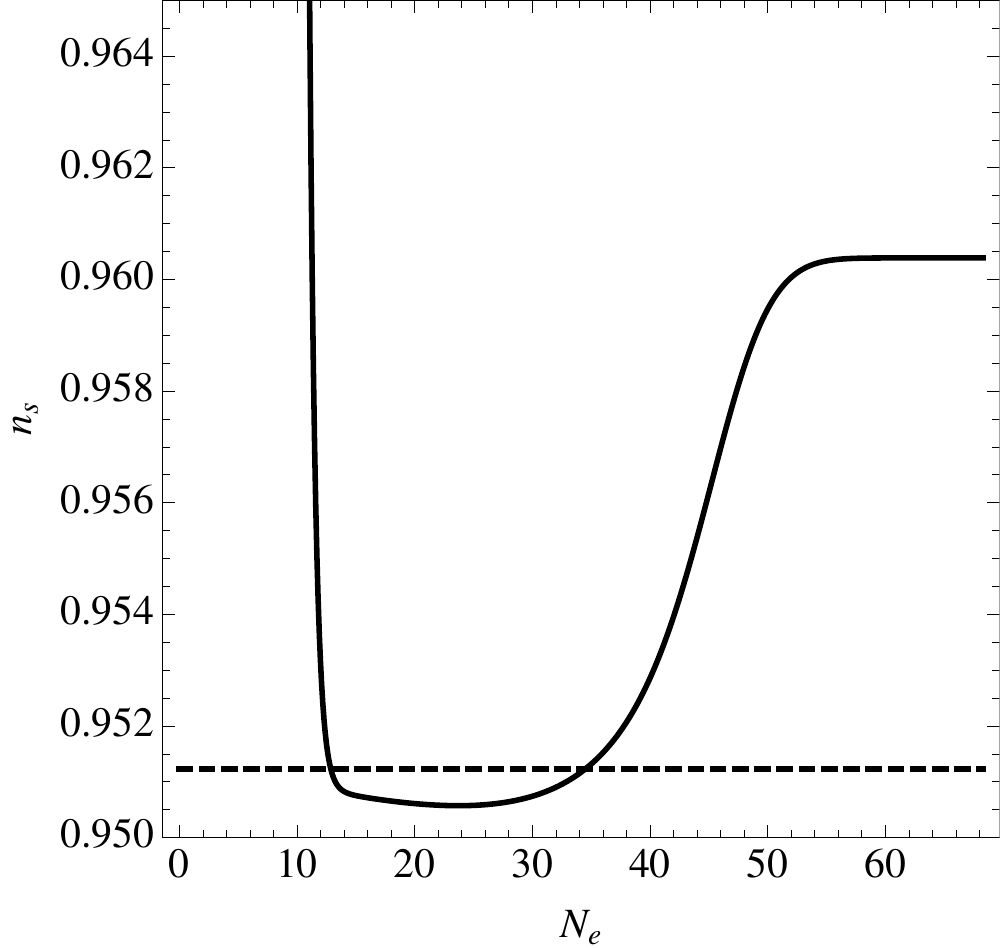}
	\end{minipage}
	\hspace{0.05cm}
\caption{Time evolution of the curvature perturbations and of the spectral index on scales $k_*=k_{60}$} for the case $A=0.83$,$\hat{G}=1$ and $\theta_0=0.9$.
	\label{fig:understanding_ns2}
\end{figure}

Starting with the tensor to scalar ratio, we observe that the effect of the multi-field dynamics is to flatten the peak and therefore to bring the results more in line with the PLANCK 2015 constraint of $r<0.114$. This effect is partially due to the  tensor modes beings unaffected by the multifield effects and in part due to the fact that the single field estimate for the amplitude of the scalar perturbations is a bad approximation for trajectories that turn early. Therefore by underestimating the amplitude of the scalar fluctuations, the single field formula overestimates the tensor to scalar ratio by
\begin{align}
r \quad &= \quad 16 \: \epsilon_* \frac{P_0}{P_{\zeta}}(k_*,\tau_{end})\ .
\label{eq: 2fr}
\end{align}

In what concerns the spectral index, we also see that the multi-field estimate is considerably sharper than what one would expect by applying single field results. We observe that the peak at low $\theta_0$ is absent and that the trough is shallower. We recall that the single field estimate for $n_s$, Eq. (\ref{eq:nsSingle}), is obtained by taking the decoupling limit $\eta_\perp\rightarrow 0$. This condition is clearly violated by the early turning trajectories, for which $\eta_\perp$ peaks as the scalar modes leave the horizon. We therefore conclude that the peak in the low $\theta_0$ region is spurious \footnote{Comparing the results for the single field estimates of figure \ref{fig:nsrA083} and those of   \cite{higgsotic} we see that they differ in the low $\theta_0$ range, where  \cite{higgsotic} has no peak. This difference can be traced back to how one generalises $\eta=\frac{V''}{V}$ for multifield cases. If one takes $\eta$ to be the smallest eigenvalue of $\frac{G^{ij}V_{ij}}{V}$ then indeed there is no peak . However we use here a different, and more accurate, prescription which can be indeed derived from the decoupling limit, as we argued in section \ref{sec:2field}.}. To understand the change in $n_s$ for larger $\theta_0$ trajectories it is useful to rewrite eq. (\ref{eq:MStau1}) as 
\begin{displaymath}
\frac{d^2 v^T_{\alpha}}{d\tau^2} +\underbrace{(\Omega_{TT}-a^2 H^2 \eta^2_{\perp}+k^2)v_{\alpha}^T}_{elastic force}=\underbrace{- 2 a H \eta_{\perp} \frac{dv_{\alpha}^N}{d\tau}-\frac{d(aH\eta_{\perp})}{d\tau}v_{\alpha}^N-\Omega_{TN}v_{\alpha}^N}_{external force}\ .
\end{displaymath}
We see that the equation of motion for the curvature perturbation is equivalent to a frictionless harmonic oscillator with a "time" dependent proper frequency  subject to an external force whose magnitude is set by the isocurvature perturbation. The effects of a turn in the background trajectory, which gives rise to the external force, will be more pronounced on $k$-modes for which the ratio between the external force and the elastic force is larger. This ratio is well approximated by the simpler relation between
\be
R(k)\simeq\frac{v_\alpha^N(k)}{v_\alpha^T(k)}\ .
\ee
Since the amplitude of different $k$-modes around the pivot scale will be affected differently by a turn in the background trajectory, there will be superhorizon evolution of the spectral index for the curvature perturbations.
From the solutions in the decoupling limit ($\eta_\perp=0$)  one finds that on superhorizon scales and before the turn
\be
v_\alpha^T\propto k^{-\nu_T}=k^{n_s^0/2-2} \qquad \text{and} \qquad v_\alpha^N\propto k^{-\nu_N}\ ,
\ee
where $n_s^0$ denotes the curvature spectral index before the turn in the trajectory and is assumed to be $n_s^0<1$. As for the isocurvature perturbations, one may expand
\be
\nu_N\sim \frac{3}{2}-\frac{1}{3} (1+2\epsilon) \left(\frac{M}{H}\right)^2 +\epsilon <\frac{3}{2} \ ,
\ee
 which leads to 
\be
R(k)\propto k^{1/2-n_s^0/2}=k^\alpha,\qquad\alpha>0 \ .
\ee
To understand how this changes the spectral index consider a pair of k-modes around the pivot scale $k_*$: $k_-<k_*<k_+$. It follows that since
\begin{displaymath}
\frac{R(k_-)}{R(k_+)}\propto \left(\frac{k_-}{k_+}\right)^{\alpha} < 1 \ ,
\end{displaymath}
the $k_+$ mode power is more enhanced than the $k_-$ mode power resulting in a spectral index closer to unity
\be
n_s^{end}>n_s^0 \ .
\ee
This behaviour can be clearly observed in figure \ref{fig:understanding_ns2}, where the $k_-$ power is less enhanced than the $k_+$, resulting in a more even distribution of power and in an increase in the spectral index. Note that, before the turn ($N_e<50$), the single field estimate is actually a good approximation to the full result and that it only fails due to the sharp turn in the background trajectory that causes conversion of isocurvature into curvature power.
\clearpage


\subsection{Modified Higg-otic regime: $A=0.7$, $\hat{G}=1$}\label{sec:A07}

The value of $A=0.83$ analysed in the previous section follows, via RG running, from the assumption that the light scalar surviving down to  the EW scale is to be identified with the SM Higgs and that no extra degrees of freedom enter the particle spectrum beyond those of the MSSM. It is however conceivable that extra light particles are present in an extension of the MSSM, modifying the running of the Higgses' masses between the compactification/string scale and the SUSY-breaking scale $M_{SS}$. If this is the case then other values of $A$ can be compatible with the Higgs-otic scenario. Keeping this possibility in mind we now analyse the case $A=0.7$,
which corresponds to a mass ratio  $m_H/m_h=2.38$ at the string scale.\\

\begin{figure}[h!]
	\centering
	\begin{minipage}[b]{0.49\linewidth}
	\centering
	\includegraphics[width=1\textwidth]{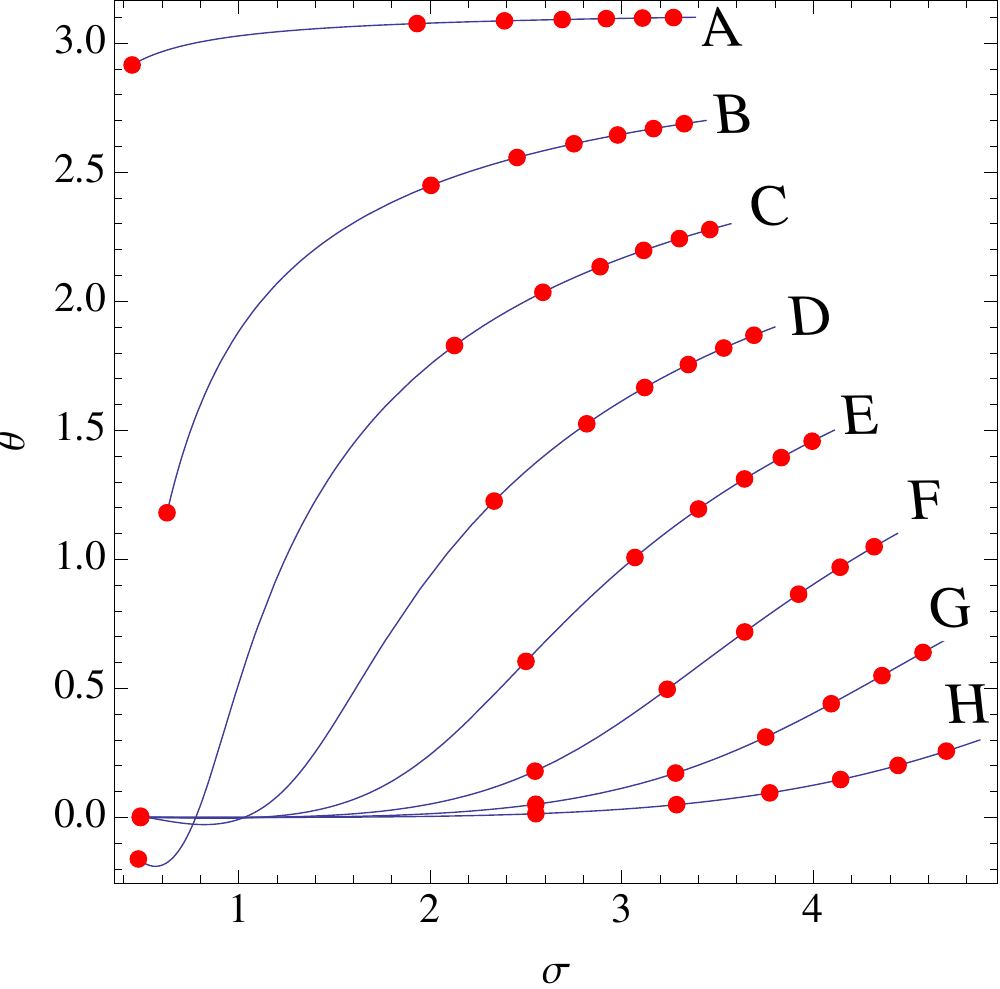}
    \end{minipage}
	\hspace{0.05cm}
	\begin{minipage}[b]{0.49\linewidth}
	\centering
\includegraphics[width=1\textwidth]{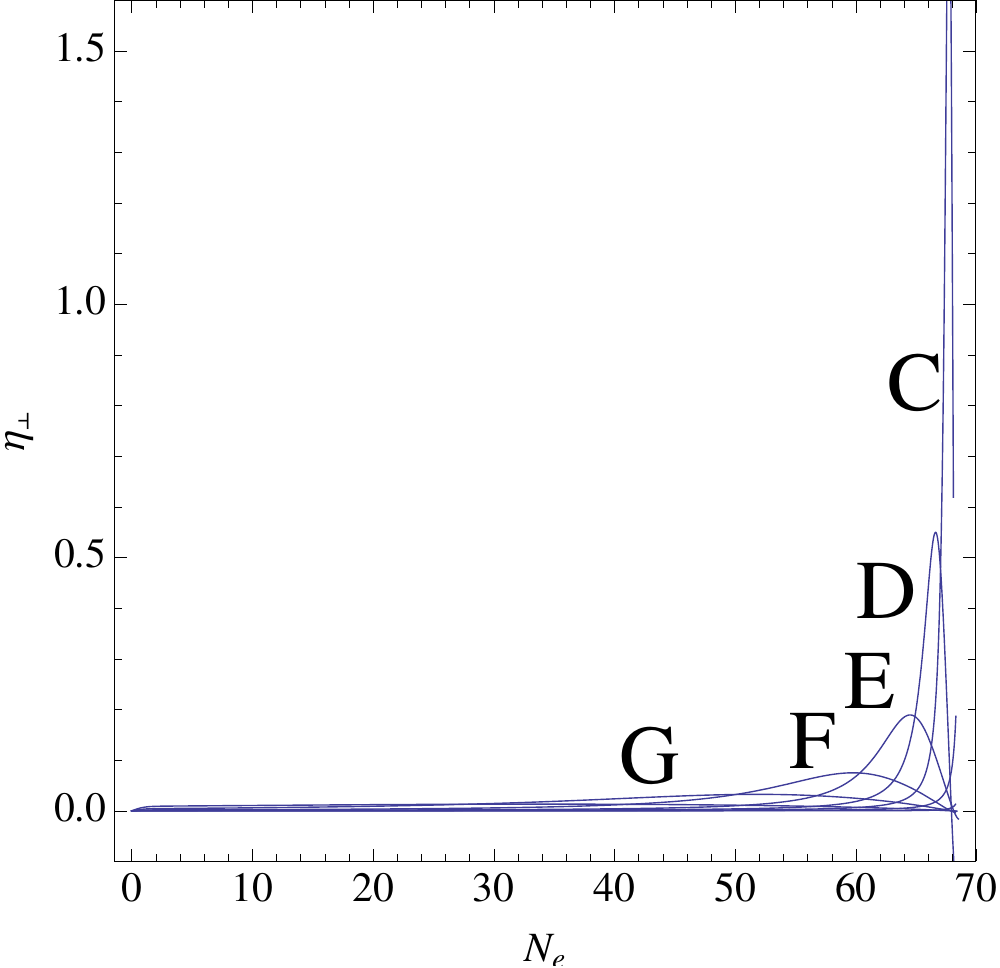}
	\end{minipage}
	\hspace{0.05cm}
\caption{Left: The last 68 efoldings for some representative inflationary trajectories. Red dots mark 10 efoldings intervals on each trajectory. Right: Evolution of $\eta_\perp$ for the different trajectories.}
	\label{fig:BGGhat1_A07}
\end{figure}

\begin{figure}[h]
	\centering
	\begin{minipage}[b]{0.49\linewidth}
	\centering
	\includegraphics[width=1\textwidth]{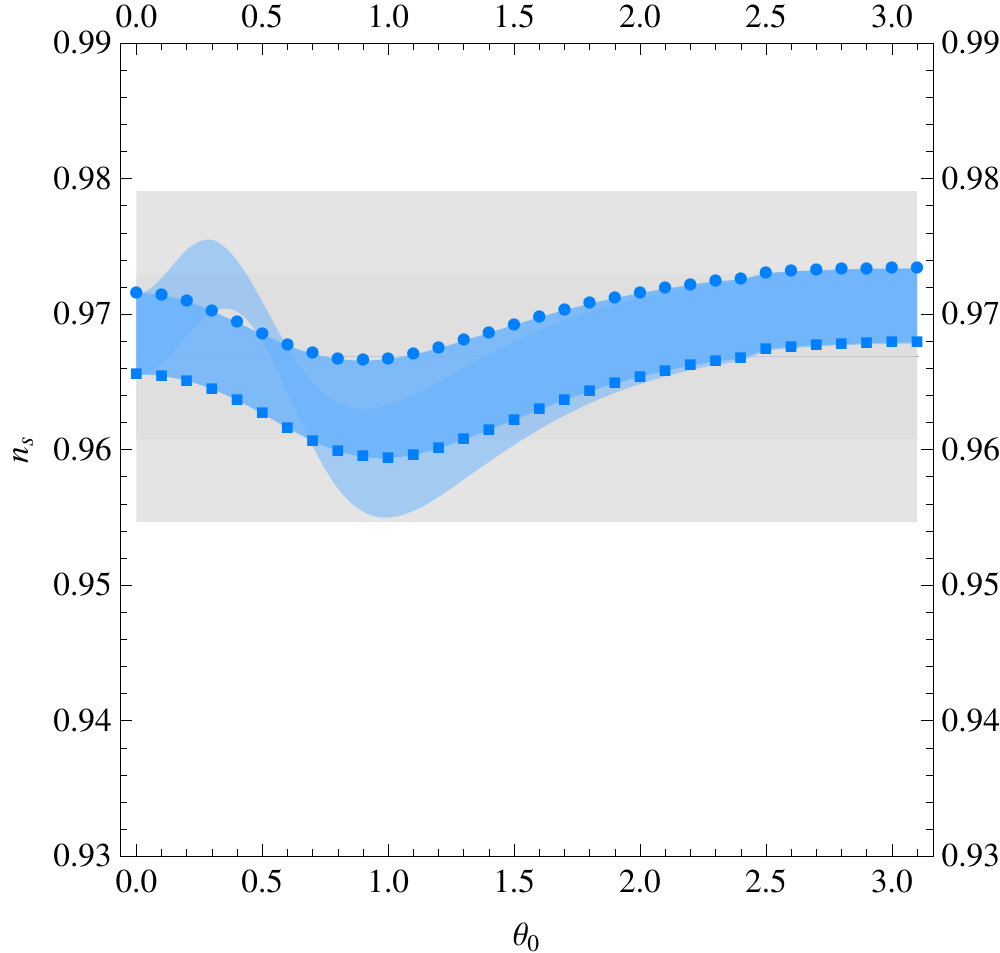}
    \end{minipage}
	\hspace{0.05cm}
	\begin{minipage}[b]{0.49\linewidth}
	\centering
\includegraphics[width=1\textwidth]{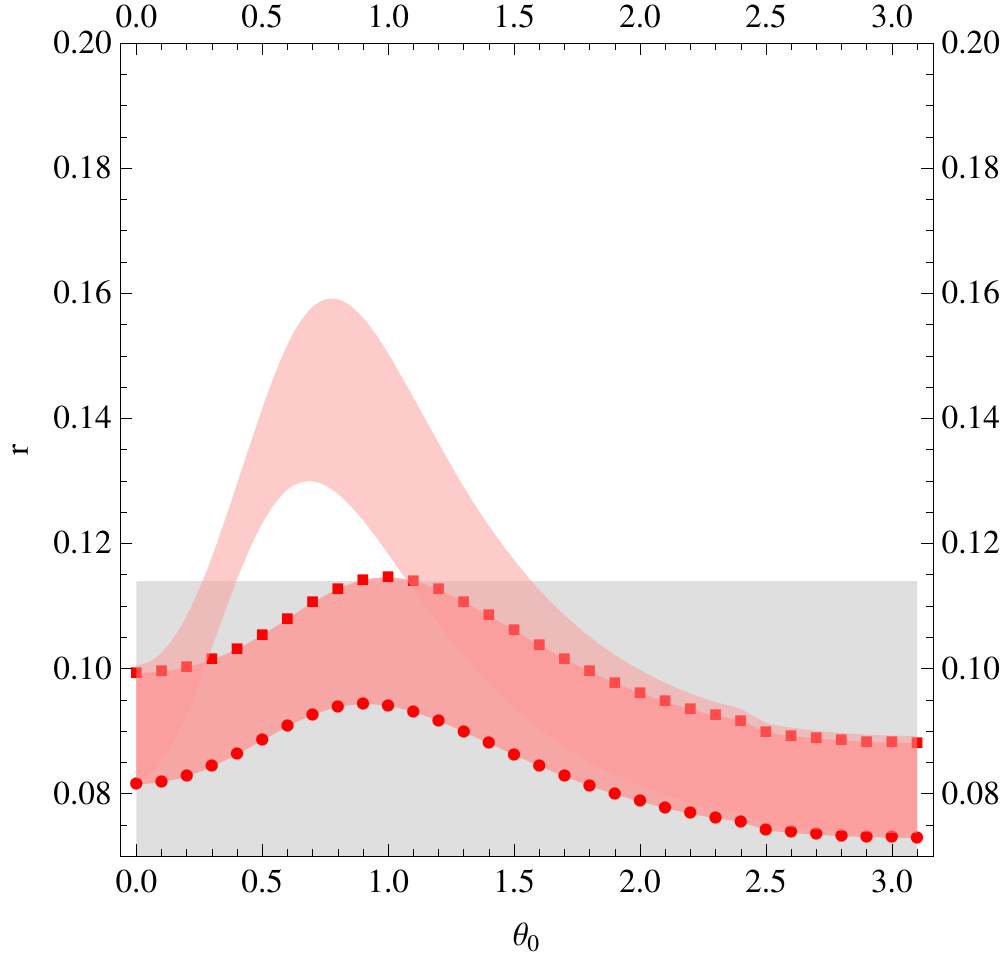}
	\end{minipage}
	\hspace{0.05cm}
\caption{Spectral index (left) and tensor to scalar ratio (right) for $\hat{G}=1$ and $A=0.7$ superimposed with the PLANCK 2015 constraints $n_s= 0.96688 \pm 0.0061$ and $r<0.114$ (grey band). The slightly transparent curve corresponds to the single field estimate while the one surrounded by the data points corresponds to the multi-field results. Circles: $N_e=60$, Squares: $N_e=50$.}
	\label{fig:nsrA07}
\end{figure}

In figure \ref{fig:BGGhat1_A07} we present sample background trajectories and the corresponding evolution of the $
\eta_\perp$ parameter.  Comparing with the results of the previous section we see that the trajectories are straighter and that the $\eta_{\perp}$ peaks are less pronounced and located at later times. This implies that the differences between the exact results and the single field estimates for the observables should be less pronounced than for the $A=0.83$ point. That is indeed the case, as can be seen from comparing figures \ref{fig:nsrA083} and \ref{fig:nsrA07}. Though less pronounced, the disparity between single field estimates and the full results is still important as the effect of the multi-field dynamics is to bring the observables more in line with the current constraints on $n_s$ and $r$: the variation in the spectral index is damped and the tensor to scalar ratio is significantly reduced. In fact $n_s$ is comfortably inside the 2$\sigma$ PLANCK 2015 band and the $r$ is reduced to the point that it complies with the observational upper bound $\forall \: \theta_0$. This highlights the importance of properly estimating the observables at a time of ever increasing measurement precision.\\

In table \ref{tab:betaIsoGhat1A07} we present the isocurvature fraction at the end of inflation as well as the comparison between the single field estimate for the adiabatic amplitude and the numerical result for the various trajectories.

\begin{table}[h]
\begin{center}
\begin{tabular}{c|c|c|c|c|c|c|c|c}
\hline
\hline
Trajectory &A&B&C&D&E&F&G&H \\
 \hline
 \hline
 $P_\zeta/P_0\big |_{k_{60}} $&1.01&  1.01&  1.02&  1.04&  1.09&  1.20& 1.40&  1.22\\
\hline
 $\log_{10} \beta_{iso}(k_{60})$ & -4& -4& -5& -9& -10& -9& -12& -13\\
 \hline
\hline
 $P_\zeta/P_0\big |_{k_{50}} $ &1.01 & 1.01& 1.02&  1.05 &  1.11&  1.26&  1.43&1.16  \\
\hline
  $\log_{10} \beta_{iso}(k_{50})$& -3 & -3& -5& -9& -9& -11& -12& -13\\
   \hline
\hline
\end{tabular}
\end{center}
\label{tab:betaIsoGhat1A07}
\caption{Order of magnitude of the isocurvature fraction at the end of inflation for the trajectories of Fig. \ref{fig:BGGhat1_A07}.}
\end{table}%


\subsection{Almost single field regime: $A=0.2$, $\hat{G}=1$}\label{sec:A02}

The Higgs-otic idea also features regimes in which the connection between inflation and MSSM Higgs physics is absent or hard to realise. In these cases, where the fluxes are such that $A$ is substantially different from the canonical value of $0.83$, inflation can still be driven by the D-brane position modulus on the compact space's $T^2$ if the
inflaton is associated to other degrees of freedom. Indeed, for small $A$ 
the values of the Higgs masses $m_h,m_H$ are too close to each other for the running from the string scale down  to the SUSY-breaking scale 
to be sufficiently strong to yield an (approximately) massless SM doublet at $M_{SS}$, but one could still identify the complex inflaton  with other 
degrees of freedom in some extension of the MSSM. Note, in particular,  that if SUSY particles are found at LHC, the
canonical Higgs-otic scenario with $A=0.83$ would be ruled out, since it assumes a large SUSY-breaking scale $M_{SS}\simeq 10^{13}$ GeV.
In this case, the inflaton could however be identified with extra scalars  which could have SUSY preserving masses at the inflaton scale 
$\simeq 10^{13}$ GeV. As we said, examples of those degrees of freedom could be right-handed doublets
of a left-right symmetric extension of the MSSM or even colour triplet scalar partners of the Higgs fields. In such a case, very small values for the
inflaton parameter $A$ would be consistent.

While the search for specific MSSM extensions 
with this structure is quite interesting,  the inflationary dynamics may be studied in a model independent manner assuming that such new 
non-Higgs degrees of freedom correspond again to the position moduli of $D$-branes.   We study here for comparison  the case with $A=0.2$ 
(corresponding to $m_H/m_h=1.22$).

\begin{figure}[h!]
	\centering
	\begin{minipage}[b]{0.49\linewidth}
	\centering
	\includegraphics[width=1\textwidth]{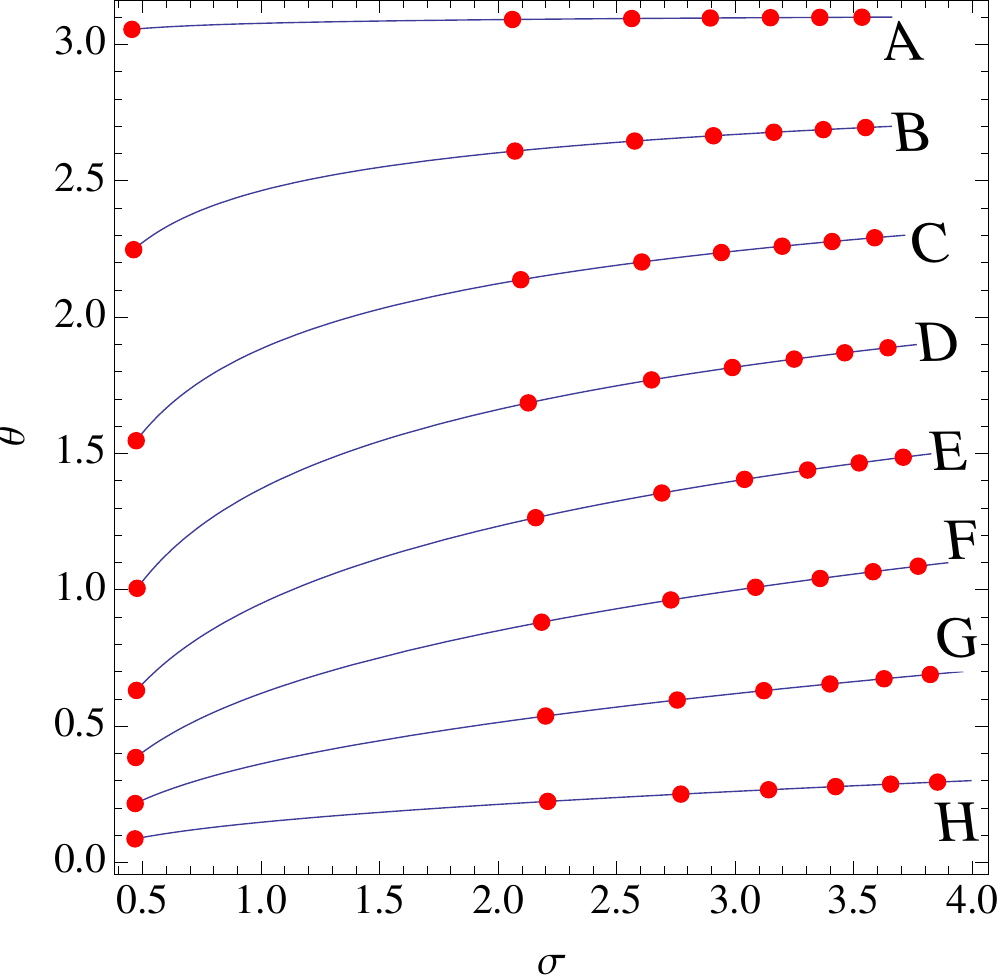}
    \end{minipage}
	\hspace{0.05cm}
	\begin{minipage}[b]{0.49\linewidth}
	\centering
\includegraphics[width=1\textwidth]{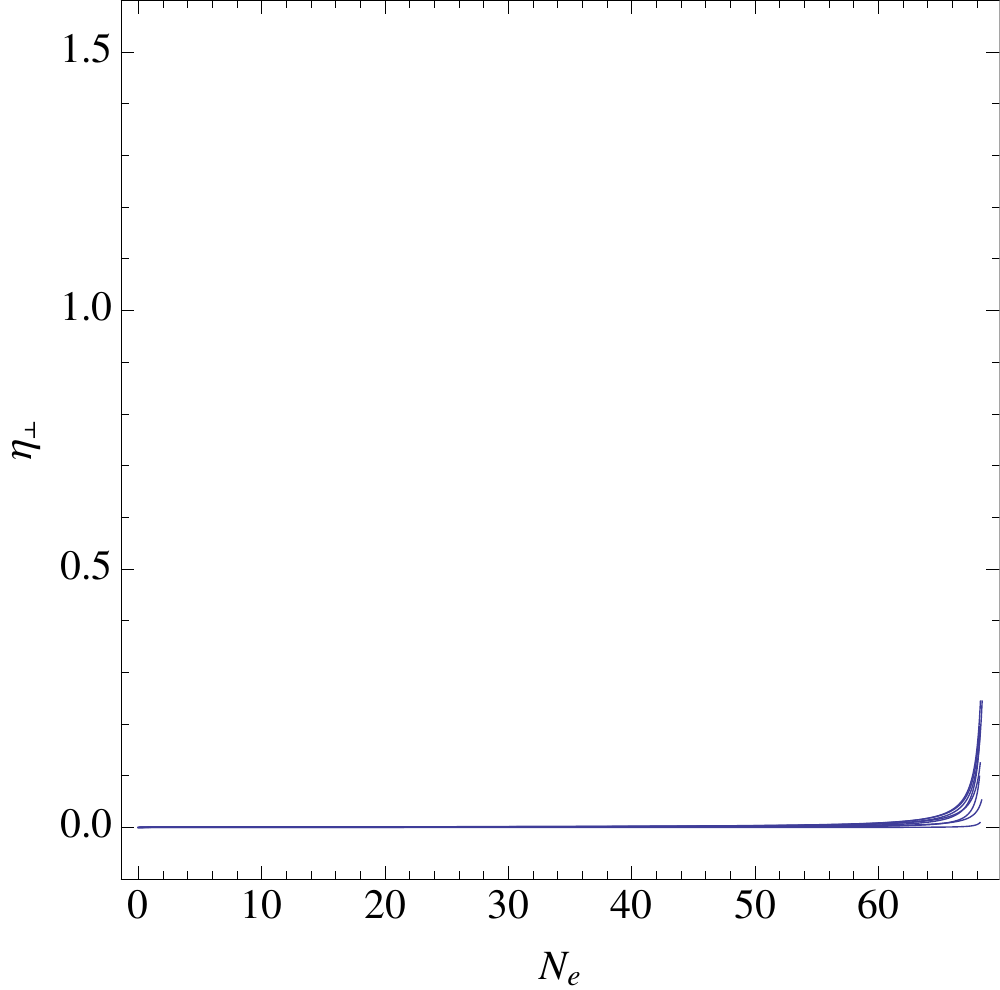}
	\end{minipage}
	\hspace{0.05cm}
\caption{Left: The last 68 efoldings for some representative inflationary trajectories. Red dots mark 10 efoldings intervals on each trajectory. Right: Evolution of $\eta_\perp$ for the different trajectories.}
	\label{fig:BGGhat1_A02}
\end{figure}

\begin{figure}[h]
	\centering
	\begin{minipage}[b]{0.49\linewidth}
	\centering
	\includegraphics[width=1\textwidth]{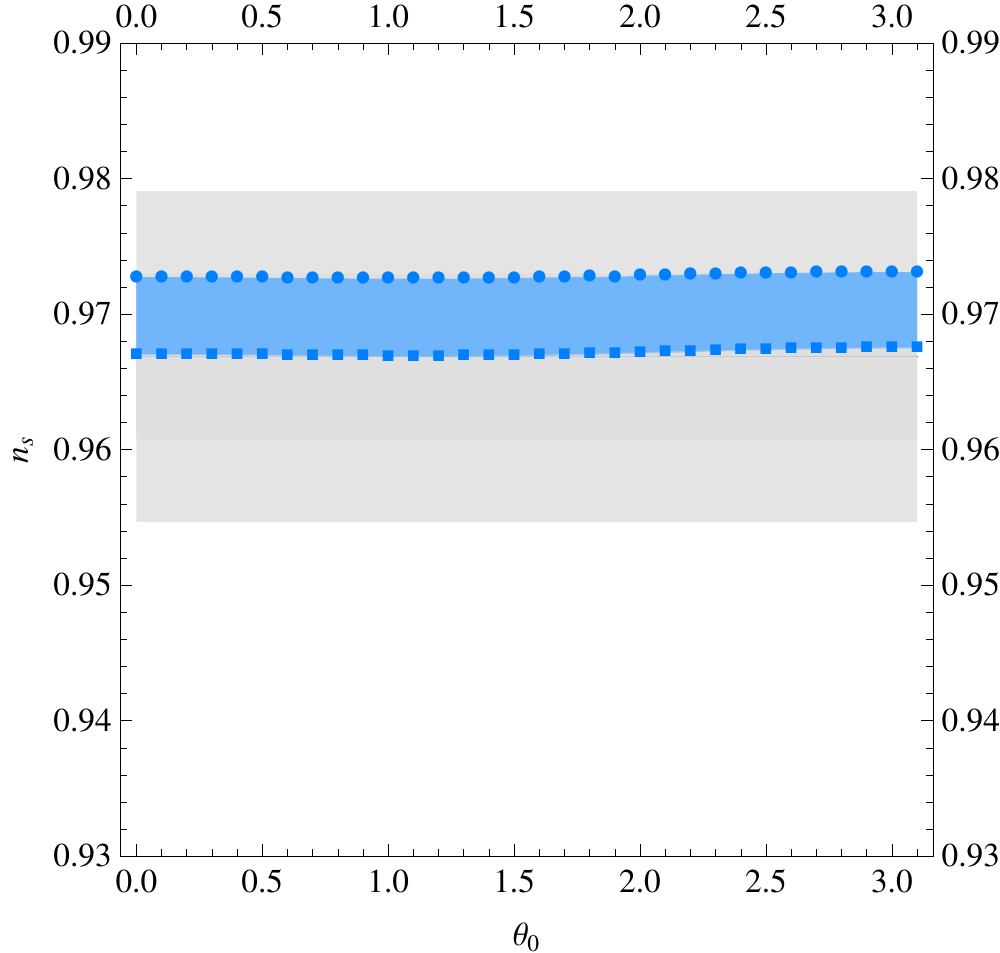}
    \end{minipage}
	\hspace{0.05cm}
	\begin{minipage}[b]{0.49\linewidth}
	\centering
\includegraphics[width=1\textwidth]{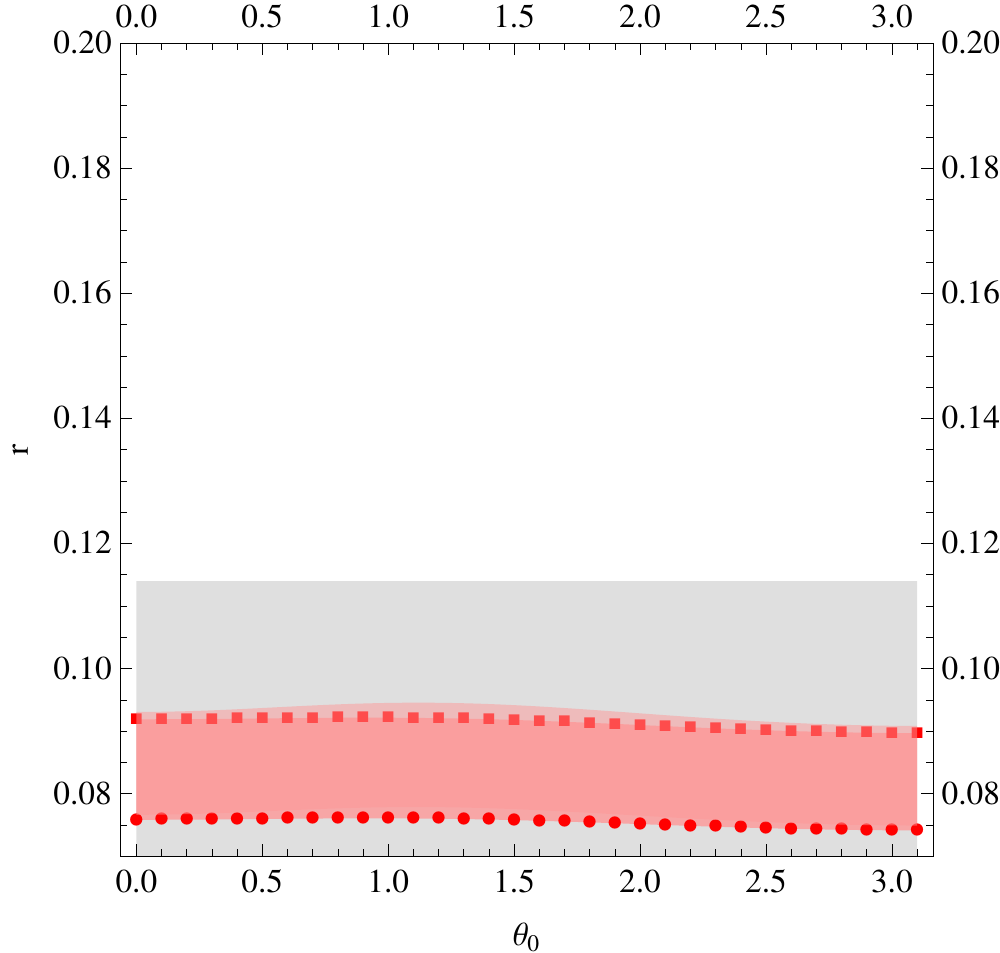}
	\end{minipage}
	\hspace{0.05cm}
\caption{Spectral index (left) and tensor to scalar ratio (right) for $\hat{G}=1$ and $A=0.2$ superimposed with the PLANCK 2015 constraints $n_s= 0.96688 \pm 0.0061$ and $r<0.114$ (grey band). The slightly transparent curve corresponds to the single field estimate while the one surrounded by the data points corresponds to the multi-field results. Circles: $N_e=60$, Squares: $N_e=50$.}
	\label{fig:nsrGhat02}
\end{figure}

In figure \ref{fig:BGGhat1_A02} we present the background evolution for sample trajectories in such a regime.
In this case we see that the trajectories are essentially straight along the $\sigma$ direction with only slight turning in the last 10 efoldings of expansion. This is in accordance with the fact that  $\theta$ becomes massless in the limit  of vanishing $A$. The straight trajectories imply that $\eta_\perp$, being inversely proportional to the curvature radius,  vanishes everywhere except at the very end of inflation (where the mild turning takes place) as can be seen in  figure \ref{fig:BGGhat1_A02}. As mentioned above $\eta_\perp$ controls the coupling between the curvature and isocurvature modes and so this limit corresponds to a decoupling regime of Eqs. \eqref{eq:MStau1} and \eqref{eq:MStau2}. We therefore expect the single field estimates of Eqs. \eqref{eq:nsSingle} and \eqref{eq:rSingle} for the inflationary observables to provide a good approximation to the full result. In fact, if one employs the multi-field formalism in the computation of the observables and compares it with the single field estimates, one finds that there is agreement at the level of a few percent. This is displayed in figure \ref{fig:nsrGhat02} where the naive single field bands track the exact results to the point of being almost indistinguishable.

Since isocurvature is practically decoupled from  curvature throughout the observable inflationary range, the isocurvature fraction at the end of inflation as estimated by $\beta_{iso}$ is larger that in the previous cases, as can be seen in table \ref{tab:betaIsoGhat1_A02}. This is due to the fact that, unable to transfer power to the adiabatic mode, all the isocurvature modes can do in their superhorizon evolution is to decay slowly. Note that even though this almost single field regime gives rise to the largest isocurvature fraction, it is still below the highest upper bound on $\beta_{iso}$ derived from the latest PLANCK 2015 data \cite{Ade:2015lrj}.\\

\begin{table}[h]
\begin{center}
\begin{tabular}{c|c|c|c|c|c|c|c|c}
\hline
\hline
Trajectory &A&B&C&D&E&F&G&H \\
 \hline
 $P_\zeta/P_0\big |_{k_{60}} $&1.01&  1.01&  1.01&  1.02&  1.02&  1.02& 1.02&  1.01\\
 \hline
  $\log_{10} \beta_{iso}(k_{60})$ & -4& -4& -5& -4& -5& -6& -6& -6\\
 \hline
 \hline
 $P_\zeta/P_0\big |_{k_{50}} $ &1.01 & 1.01& 1.02&  1.02 &  1.03&  1.03&  1.02&1.02  \\
\hline
 $\log_{10} \beta_{iso}(k_{50})$& -4 & -4& -4& -4& -5& -5& -6& -6\\
 \hline
\hline
\end{tabular}
\end{center}
\label{tab:betaIsoGhat1_A02}
\caption{Ratio between the amplitude of the curvature perturbations at the end of inflation and the single field estimate of eq. \eqref{eq:nsSingle} and amplitude of isocurvature perturbations for the trajectories of fig. \ref{fig:BGGhat1_A02}.}
\end{table}%

\subsection{Varying $\hat{G}$}

\begin{figure}[h]
\begin{center}
\includegraphics[width=0.7\textwidth]{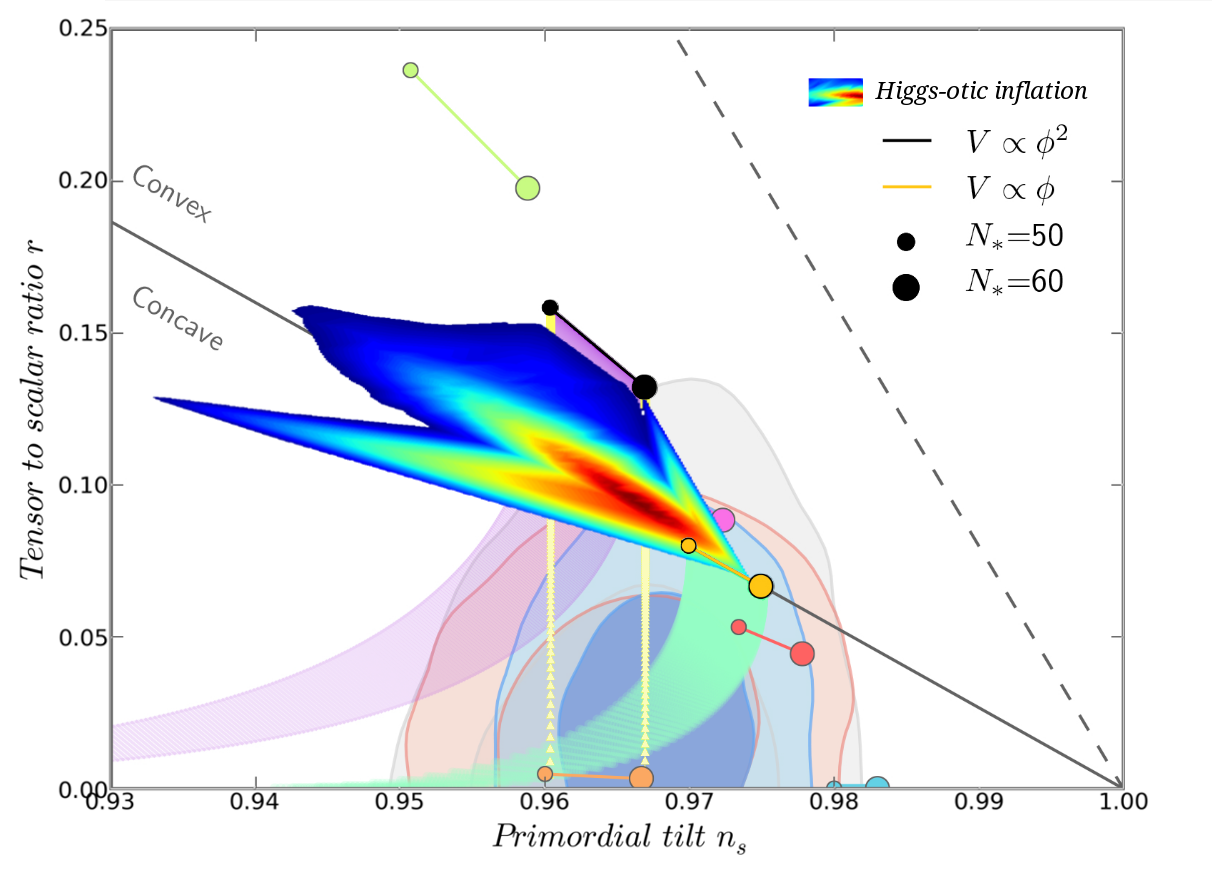}
\caption{Tensor to scalar ratio and spectral index for Higgs-otic inflation with $A=0.83$ and arbitrary $\hat G$. The data is superimposed over the recent Planck exclusion limits \cite{Ade:2015lrj}. The color pattern (from red to blue) corresponds to (higher or lower) density of initial condition points . There is a lower cutoff on the density required to be plotted (or equivalently in the level of fine-tuning allowed) missing  around 10 \% of the points.}
\label{rns_G}
\end{center}
\end{figure}

Finally let us discuss the effects of varying $\hat{G}$ over the results. As we commented in section \ref{sec:Higgsotic}, $\hat G$ parametrises the total amount of flux quanta in Planck units, which determines in turn the ratio between the supersymmetry breaking scale and the string scale. In particular,
\beq
\hat G=(V_4\mu_7)^{-1/2}G_3\simeq \frac{n}{M_p}\sim 5\frac{M_{SS}}{M_s^2}
\eeq
with $n$ being the flux quanta. We will take the string scale of order $M_s\sim 10^{16}-10^{17}$ GeV, as suggested by gauge unification. An isotropic compactification with $n\sim \mathcal{O}(1)$ then implies $M_{SS}\sim 10^{13}$ GeV. But it should be noticed that the parameter $n$ in general receives contributions from a large number of 3-cycles so that large cancellations can take place that lead to $n\ll 1$,  lowering the scale of
SUSY breaking. However, $n\ll 1$ is problematic if one wants to satisfy the experimental bounds on density scalar perturbations. A lower supersymmetry breaking scale may lead to a too low amplitude of the scalar power spectra in this inflationary model. The best fit indeed corresponds to $M_{SS}\sim 10^{13}$, which is the typical scale for flux-induced supersymmetry breaking obtained by assuming the flux quanta to be of order one. On the other hand, $n\gg 1$ (and thereby a higher supersymmetry breaking scale) makes the potential energy trans-Planckian at the beginning of inflation, which is inconsistent with the effective field theory approach. Throughout this paper we have shown the results for the representative value   $\hat G=1$ in Planck units. Here we also show the results for  other possible values of $\hat G$ for completeness. In figure \ref{rns_G} we plot our results in the $r-n_s$ plane superimposed over the experimental Planck exclusion limits. The data correspond to $A=0.83$ and arbitrary $\hat G$. 

Notice that $\hat G$ not only enters in the absolute value of the potential, but also in the field redefinition to get canonical kinetic terms. The bigger $\hat G$ is, the stronger is the flattening of the potential and the results are closer to those of linear inflation. This is the reason why  our results actually interpolate between quadratic (small $\hat G$, negligible flattening) and linear (big $\hat G$, strong flattening). However, as we already said, those closer to linear inflation are in better agreement with both density scalar perturbations and tensor-to-scalar ratio constraints from Planck. In figure \ref{rns_G} the color pattern from red to blue refers to the density of points, being the red regions the most populated. This could have been anticipated from figure \ref{fig:nsrA083}, where  most of the initial conditions gave rise to values of $r$ and $n_s$ closer to the single field prediction. The spreading of the results to smaller values of the spectral index is due to the freedom on the choice of the initial conditions, but as we said the blue region corresponds to very fine-tuned values of $\theta_0$ and the majority of the points is localised at $n_s\simeq 0.965$ and $r\simeq 0.08-0.12$. It is remarkable that the most populated regions are indeed those in better agreement with experiments.

\section{Conclusions}\label{sec:conclusions}

The current lack of experimental evidence for low-energy supersymmetry coupled with the possibility of observing primordial gravitational waves in the near future drives us to analyse a model in which these two seemingly unconnected phenomena have their roots in the same sector of the theory: Higgs-otic inflation.

In Higgs-otic inflation the scale of SUSY breaking, $M_{SS}\sim 10^{13}$ GeV, is identified with the inflation's mass and the Universe's expansion is driven by the MSSM Higgs scalars. Such a high inflationary scale gives rise to a sizeable tensor fraction and implies super-Planckian field excursions for the inflatons. The model is embedded into a local string construction, in which the Higgses/inflatons are D-brane position moduli. Such a stringy embedding allows for control over higher order corrections to the inflationary potential that are usually problematic for large field inflation models.

In this work we have analysed in detail the observational signatures of this model, taking into account the effects of isocurvature perturbations that are unavoidable whenever there are multiple dynamical fields during inflation. We found that the effects of 2-field dynamics on the observables are more pronounced whenever there is a turn in the background trajectory. In Higgs-otic inflation the presence of such a turn during the last 60 efoldings of expansion is controlled by the choice of background fluxes. For flux choices leading to almost straight trajectories, the single field estimates for the inflationary observables derived in \cite{higgsotic} are a good approximation to the full result as we have seen in Sec. \ref{sec:A02}. 
On the other hand for those flux choices consistent with the 
identification of the inflaton with the MSSM Higgses, we find that 
the trajectories will necessarily present some degree of turning.  At the turn isocurvature and curvature perturbations are coupled, with isocurvature sourcing the curvature power. This implies superhorizon evolution of the curvature perturbations and translates into an increase of the adiabatic perturbations' amplitude and the corresponding decrease in the tensor to scalar ratio. Since different $k$-modes are affected differently by a turn in the trajectory, there will be superhorizon evolution of the spectral index, which we have shown to lead to a sharper prediction for $n_s$ when compared to the single field approximation employed in \cite{higgsotic}. In what concerns the isocurvature power at the end of inflation, we have shown that it is always suppressed enough to be compatible with the less stringent upper bounds from PLANCK 2015.  We have shown in \ref{sec:A083} and \ref{sec:A07} that the observational footprint of Higgs-otic inflation is narrower than what one would expect by simply performing an adiabatic projection: the range over which $n_s$ varies is smaller and centred around the Planck best fit value and the tensor to scalar ratio is reduced significantly  yielding a region $r=0.08-0.12$ depending on the model initial conditions.

In summary, our analysis demonstrates that the observable signatures of Higgs-otic inflation are in line with the results of the  joint Planck/BICEP analysis \cite{Ade:2015tva}. It would be interesting to address additional questions like moduli stabilisation and reheating in this model. We leave those aspects to future research.

\vspace{1cm}

\noindent 
\centerline
{\bf Acknowledgements}
\\

We would like to thank F. Marchesano,  A.~Uranga  and A. Westphal for useful discussions. We acknowledge the use of \cite{Dias:2015rca,Price:2014xpa,Easther:2013rva,Lalak:2007vi,Turzynski:2014tza,Avgoustidis:2011em,Huston:2011fr} in the development and debugging of our numerical methods. This work has been supported by the ERC Advanced Grant SPLE under contract ERC-2012-ADG-20120216-320421, by the grant FPA2012-32828 from the MINECO,  and the grant SEV-2012-0249 of the ``Centro de Excelencia Severo Ochoa" Programme.  I.V. is supported through the FPU grant AP-2012-2690. 

\end{document}